\documentclass[journal]{IEEEtran}

\usepackage{epsfig} 
\usepackage{times} 
\usepackage{amsmath} 
\usepackage{amssymb}  
\usepackage{amsthm}
\usepackage{algorithm,algorithmic}
\usepackage{multirow}
\usepackage{array}
\usepackage{cite}
\usepackage{color}
\usepackage{url}

\begin{document}

\title{Optimization-based Control for Bearing-only Target Search with a Mobile Vehicle}

\author{Zhuo Li, Keyou You, Shiji Song, Anke Xue
\thanks{This work was supported by National Key  Research and Development Program of China under Grant No.2016YFC0300801, and National Natural Science Foundation of China under Grant No.41576101.}
\thanks{Z. Li, K. You and S. Song are with Department of Automation and BNRist, Tsinghua University, Beijing 100084, China (youky@tsinghua.edu.cn). A. Xue is with the Key Laboratory for IOT  and Information Fusion Technology of Zhejiang, Institute of Information and Control, Hangzhou Dianzi University, Hangzhou 310018, China (e-mail: akxue@hdu.edu.cn). }}%


\def\inftyn #1{\|#1\|_{\infty}}
\def\twon #1{\|#1\|}
\def\rainfty{\rightarrow\infty}
\def\ra{\rightarrow}
\def\red{\textcolor{red}}
\def\rk{\text{rank}}
\def\argmin{\text{argmin}}
\def\argmax{\text{argmax}}
\def\bA{\mathbb{A}}
\def\bB{\mathbb{B}}
\def\bC{\mathbb{C}}
\def\bD{\mathbb{D}}
\def\bE{\mathbb{E}}
\def\bF{\mathbb{F}}
\def\bG{\mathbb{G}}
\def\bH{\mathbb{H}}
\def\bI{\mathbb{I}}
\def\bJ{\mathbb{J}}
\def\bK{\mathbb{K}}
\def\bL{\mathbb{L}}
\def\bM{\mathbb{M}}
\def\bN{\mathbb{N}}
\def\bO{\mathbb{O}}
\def\bP{\mathbb{P}}
\def\cQ{\mathbb{Q}}
\def\bR{\mathbb{R}}
\def\bS{\mathbb{S}}
\def\bT{\mathbb{T}}
\def\bU{\mathbb{U}}
\def\bV{\mathbb{V}}
\def\bW{\mathbb{W}}
\def\bX{\mathbb{X}}
\def\bY{\mathbb{Y}}
\def\bZ{\mathbb{Z}}

\def\cA{\mathcal{A}}
\def\cB{\mathcal{B}}
\def\cC{\mathcal{C}}
\def\cD{\mathcal{D}}
\def\cE{\mathcal{E}}
\def\cF{\mathcal{F}}
\def\cG{\mathcal{G}}
\def\cH{\mathcal{H}}
\def\cI{\mathcal{I}}
\def\cJ{\mathcal{J}}
\def\cK{\mathcal{K}}
\def\cL{\mathcal{L}}
\def\cM{\mathcal{M}}
\def\cN{\mathcal{N}}
\def\cO{\mathcal{O}}
\def\cP{\mathcal{P}}
\def\cQ{\mathcal{Q}}
\def\cR{\mathcal{R}}
\def\cS{\mathcal{S}}
\def\cT{\mathcal{T}}
\def\cU{\mathcal{U}}
\def\cV{\mathcal{V}}
\def\cW{\mathcal{W}}
\def\cX{\mathcal{X}}
\def\cY{\mathcal{Y}}
\def\cZ{\mathcal{Z}}

\def \tilA{\widetilde{A}}
\def \tilB{\widetilde{B}}
\def \tilC{\widetilde{C}}
\def \tilD{\widetilde{D}}
\def \tilE{\widetilde{E}}
\def \tilF{\widetilde{F}}
\def \tilG{\widetilde{G}}
\def \tilH{\widetilde{H}}
\def \tilI{\widetilde{I}}
\def \tilJ{\widetilde{J}}
\def \tilK{\widetilde{K}}
\def \tilL{\widetilde{L}}
\def \tilM{\widetilde{M}}
\def \tilN{\widetilde{N}}
\def \tilO{\widetilde{O}}
\def \tilP{\widetilde{P}}
\def \tilQ{\widetilde{Q}}
\def \tilR{\widetilde{R}}
\def \tilS{\widetilde{S}}
\def \tilT{\widetilde{T}}
\def \tilU{\widetilde{U}}
\def \tilV{\widetilde{V}}
\def \tilW{\widetilde{W}}
\def \tilX{\widetilde{X}}
\def \tilY{\widetilde{Y}}
\def \tilZ{\widetilde{Z}}

\def \haA{\widehat{A}}
\def \haB{\widehat{B}}
\def \haC{\widehat{C}}
\def \haD{\widehat{D}}
\def \haE{\widehat{E}}
\def \haF{\widehat{F}}
\def \haG{\widehat{G}}
\def \haH{\widehat{H}}
\def \haI{\widehat{I}}
\def \haJ{\widehat{J}}
\def \haK{\widehat{K}}
\def \haL{\widehat{L}}
\def \haM{\widehat{M}}
\def \haN{\widehat{N}}
\def \haO{\widehat{O}}
\def \haP{\widehat{P}}
\def \haQ{\widehat{Q}}
\def \haR{\widehat{R}}
\def \haS{\widehat{S}}
\def \hat{\widehat{T}}
\def \haU{\widehat{U}}
\def \haV{\widehat{V}}
\def \haW{\widehat{W}}
\def \haX{\widehat{X}}
\def \haY{\widehat{Y}}
\def \haZ{\widehat{Z}}

\def \habfA{\widehat{\bf A}}
\def \habfB{\widehat{\bf B}}
\def \habfC{\widehat{\bf C}}
\def \habfD{\widehat{\bf D}}
\def \habfE{\widehat{\bf E}}
\def \habfF{\widehat{\bf F}}
\def \habfG{\widehat{\bf G}}
\def \habfH{\widehat{\bf H}}
\def \habfI{\widehat{\bf I}}
\def \habfJ{\widehat{\bf J}}
\def \habfK{\widehat{\bf K}}
\def \habfL{\widehat{\bf L}}
\def \habfM{\widehat{\bf M}}
\def \habfN{\widehat{\bf N}}
\def \habfO{\widehat{\bf O}}
\def \habfP{\widehat{\bf P}}
\def \habfQ{\widehat{\bf Q}}
\def \habfR{\widehat{\bf R}}
\def \habfS{\widehat{\bf S}}
\def \habfT{\widehat{\bf T}}
\def \habfU{\widehat{\bf U}}
\def \habfV{\widehat{\bf V}}
\def \habfW{\widehat{\bf W}}
\def \habfX{\widehat{\bf X}}
\def \habfY{\widehat{\bf Y}}
\def \habfZ{\widehat{\bf Z}}

\def \bfA{{\bf A}}
\def \bfB{{\bf B}}
\def \bfC{{\bf C}}
\def \bfD{{\bf D}}
\def \bfE{{\bf E}}
\def \bfF{{\bf F}}
\def \bfG{{\bf G}}
\def \bfH{{\bf H}}
\def \bfI{{\bf I}}
\def \bfJ{{\bf J}}
\def \bfK{{\bf K}}
\def \bfL{{\bf L}}
\def \bfM{{\bf M}}
\def \bfN{{\bf N}}
\def \bfO{{\bf O}}
\def \bfP{{\bf P}}
\def \bfQ{{\bf Q}}
\def \bfR{{\bf R}}
\def \bfS{{\bf S}}
\def \bfT{{\bf T}}
\def \bfU{{\bf U}}
\def \bfV{{\bf V}}
\def \bfW{{\bf W}}
\def \bfX{{\bf X}}
\def \bfY{{\bf Y}}
\def \bfZ{{\bf Z}}

\def \bfa{{\bf a}}
\def \bfb{{\bf b}}
\def \bfc{{\bf d}}
\def \bfd{{\bf d}}
\def \bfe{{\bf e}}
\def \bff{{\bf f}}
\def \bfg{{\bf g}}
\def \bfh{{\bf h}}
\def \bfi{{\bf i}}
\def \bfj{{\bf j}}
\def \bfk{{\bf k}}
\def \bfl{{\bf l}}
\def \bfm{{\bf m}}
\def \bfn{{\bf n}}
\def \bfo{{\bf o}}
\def \bfp{{\bf p}}
\def \bfq{{\bf q}}
\def \bfr{{\bf r}}
\def \bfs{{\bf s}}
\def \bft{{\bf t}}
\def \bfu{{\bf u}}
\def \bfv{{\bf v}}
\def \bfw{{\bf w}}
\def \bfx{{\bf x}}
\def \bfy{{\bf y}}
\def \bfz{{\bf z}}

\def\diag{\text{diag}}
\def\tr{\text{tr}}
\def\sin{\text{sin}}
\def\cos{\text{cos}}

\def \qed {\hfill \vrule height6pt width 6pt depth 0pt}
\def\bee{\begin{equation}}
\def\ene{\end{equation}}
\def\beq{\begin{eqnarray}}
\def\enq{\end{eqnarray}}
\def\as{\overset{\text{{\em a.e.}}}{\ra}}
\newcommand{\BOX}{\hfill\rule{2mm}{2mm}}
\newtheorem{defi}{Definition}
\newtheorem{exmp}{Example}
\newtheorem{assum}{Assumption}
\newtheorem{thm}{Theorem}
\newtheorem{lemma}{Lemma}
\newtheorem{rem}{Remark}
\newtheorem{prop}{proposition}
\theoremstyle{remark}
\newtheorem*{prf}{Proof}
\newtheorem*{notation}{Notation}
\renewcommand{\algorithmicrequire}{\textbf{Input:}}
\renewcommand{\algorithmicensure}{\textbf{Output:}}
\renewcommand{\algorithmicrepeat}{\textbf{Repeat:}}
\renewcommand{\algorithmicwhile}{\textbf{While}}

\def\bh{{\bf h}}
\def\bone{{\bf 1}}
\def\bth{{\bf \theta}}
\def\hth{\widehat{\bf \theta}}
\def\tth{\widetilde{\bf \theta}}
\def\blue{\textcolor{blue}}

\maketitle

\begin{abstract}
This work aims to design an optimization-based controller for a discrete-time Dubins vehicle to approach a target with unknown position as fast as possible by only using bearing measurements. To this end, we propose a bi-objective optimization problem, which jointly considers the  performance of estimating the unknown target position and controlling the mobile vehicle to a known position, and then adopt a weighted sum method with normalization to solve it. The controller is given based on the solution of the optimization problem in ties with a least-square estimate of the target position. Moreover, the controller does not need the vehicle's global position information. Finally, simulation results are included to validate the effectiveness of the proposed controller.
\end{abstract}


\IEEEpeerreviewmaketitle

\section{Introduction}

\IEEEPARstart{T}{arget} search is a problem of steering single or multiple mobile vehicles to a target with unknown position. It has a broad range of applications, such as search and rescue in disaster response settings \cite{chen2015optimal}, planetary and undersea exploration \cite{raja2015new, best2017path}, search for lost targets \cite{lanillos2012minimum}, and environmental monitoring \cite{song2018e}. In these scenarios, it is significant and challenging for mobile vehicles to reach target positions as fast as possible, which is the motivation of this work.

If the target position is known, quite a few approaches have been proposed to minimize the searching time, see \cite{ding2010multi, zhang2012planning, otte2013c, koziol2017single} and references therein. Clearly, these approaches cannot be directly applied here. To estimate unknown target positions, various sensor measurements have been utilized, such as bearing angles \cite{bishop2007optimality, mavrommati2018real}, ranges \cite{moreno2013optimal}, and time difference of arrivals \cite{dogancay2012uav, meng2016communication}. However, most of existing works only consider either controlling a mobile vehicle to a known target position as fast as possible \cite{ding2010multi, zhang2012planning, otte2013c, koziol2017single} or maximizing the estimation performance of sensor measurements \cite{bishop2007optimality, mavrommati2018real, moreno2013optimal, dogancay2012uav, meng2016communication, lorussi2001optimal, hinson2013path, cognetti2018optimal}. In this work, they are called the control objective and the estimation objective, respectively, and are both essential for the target search problem. Based on these observations, a naive idea is to separately study these two objectives \cite{deghat2014localization, guler2017adaptive, liu2018a}, i.e.,  directly control the vehicle  to the estimated target position.

Unfortunately, a reliable estimate cannot always be obtained, particularly in the initial stage of the target search task where sensor measurements are limited. Clearly, it is not necessary to control the vehicle to such an unreliable estimated position. In fact, the controller should depend on the quality of the target  position estimate, which is again significantly affected by the controlled trajectory of the mobile vehicle. That is, the control and the estimation objectives should be jointly considered. To achieve it, we propose a bi-objective optimization problem subject to the dynamical constraint of a discrete-time Dubins vehicle, where the first objective is to control the target as fast as possible to the target (assuming it  is known) and the second objective is to optimize the estimation performance of the unknown target position. 

Usually, the vehicle cannot achieve the control and the estimation objectives simultaneously. Note that the trajectory for the control objective is a line segment from its initial position pointing at the true target, which potentially leads to a poor estimate of the target position \cite{bishop2010optimality}. However, the trajectory for the estimation objective significantly deviates from the line segment. Thus, it is essential to balance the two objectives. This work adopts a weighted sum method \cite{marler2010weighted} to transform the bi-objective optimization problem into a composite optimization problem with a weighting parameter representing the relative importance between the two objectives. Although this idea has been widely adopted in \cite{ergezer2013path, duan2014pigeon, sidoti2017a, maurovi2018path}, it is still difficult to select an appropriate weighting parameter due to different physical meanings of the objective functions. For example, the objective functions in \cite{maurovi2018path} represent the angle difference and the distance difference, which are also of different magnitudes. To this end, we rewrite the composite objective problem by normalizing the two objective functions and introducing a new weighting parameter. It proves to be much easier to select an appropriate value of the new weighting parameter, since both objective functions are normalized to be dimensionless and between zero and one.

However, the normalization always results in a complicated form of the composite objective function. Furthermore, the discrete-time Dubins vehicle \cite{liu2015discrete} adopted in this work brings a challenging constraint in the composite optimization problem, which is different from the single integrators in \cite{dogancay2012uav, meng2016communication, deghat2014localization, guler2017adaptive}. In the literature, evolutionary algorithms are commonly-used, such as the genetic algorithm in \cite{ergezer2013path}, pigeon inspired optimization method in \cite{duan2014pigeon}, and value iteration-like algorithm in \cite{Candido2011Minimum}. Moreover, the graph-based searching method is thoroughly investigated in \cite{sidoti2017a, maurovi2018path}. Different from the aforementioned approaches, we attack the complicated optimization problem by decomposing the forward velocity of the vehicle and then transforming the optimization problem into a solvable form. Particularly, an explicit solution can be directly obtained for some weighting parameters with this approach.

Then, the controller for target search can be given in terms of the solution to the optimization problem and the vehicle's global position information. Since the solution to the optimization problem depends on the unknown target position, the controller cannot be directly implemented. To solve this problem, we design a recursive nonlinear least-square estimate \cite{gavish1992performance} of the target position for the optimization-based controller by utilizing the certainty-equivalence principle. This substantially distinguishes our controller from those in \cite{deghat2014localization, guler2017adaptive, liu2018a}. 

In addition, the vehicle's global position is not always available in GPS-denied environments, although it is used in \cite{dogancay2012uav, meng2016communication, lorussi2001optimal, hinson2013path, cognetti2018optimal, deghat2014localization, guler2017adaptive, liu2018a}. In this work, we further propose a controller without the vehicle's global position. Observe that the vehicle only aims to reach the target position, for which the relative position information between the vehicle and the target is sufficient. Therefore, we utilize the bearing angles relative to the target in the vehicle's own coordinate, and derive an optimization-based controller for the global position system (GPS) -denied target search problem together with the vehicle's orientation angles from a compass. Similarly, the GPS information is not required in \cite{lin2016distributed, zheng2015enclosing} to estimate the target position, whereas they employ a group of vehicles to measure the bearing angles relative to their neighbors and to communicate individual estimates with their neighbors. In particular, the communication graphs have to be connected. 

The rest of this paper is organized as follows. In Section \ref{sec:formulation}, the target search problem is formulated as a bi-objective optimization problem for a discrete-time Dubins vehicle. In Section \ref{sec:reformulation}, we adopt the weighted sum method to solve the bi-objective optimization problem, which is rewritten by transformation and normalization. The optimization-based controller design is given in Section \ref{sec:controller}, with a recursive estimator and an optimization solver. Section \ref{sec:controller2} considers the target search problem in GPS-denied environments, where the controller is based on the bearing angles in the vehicle's coordinate. Simulations are conducted in Section \ref{sec:simulations} to illustrate the effectiveness of the proposed controllers. Finally, some conclusion remarks are drawn in Section \ref{sec:conclusions}.

\section{Problem formulation} 
\label{sec:formulation}

\subsection{The target search problem with a mobile vehicle}
\label{subsec:prob}

\begin{figure}
  \centering
\includegraphics[width=8cm]{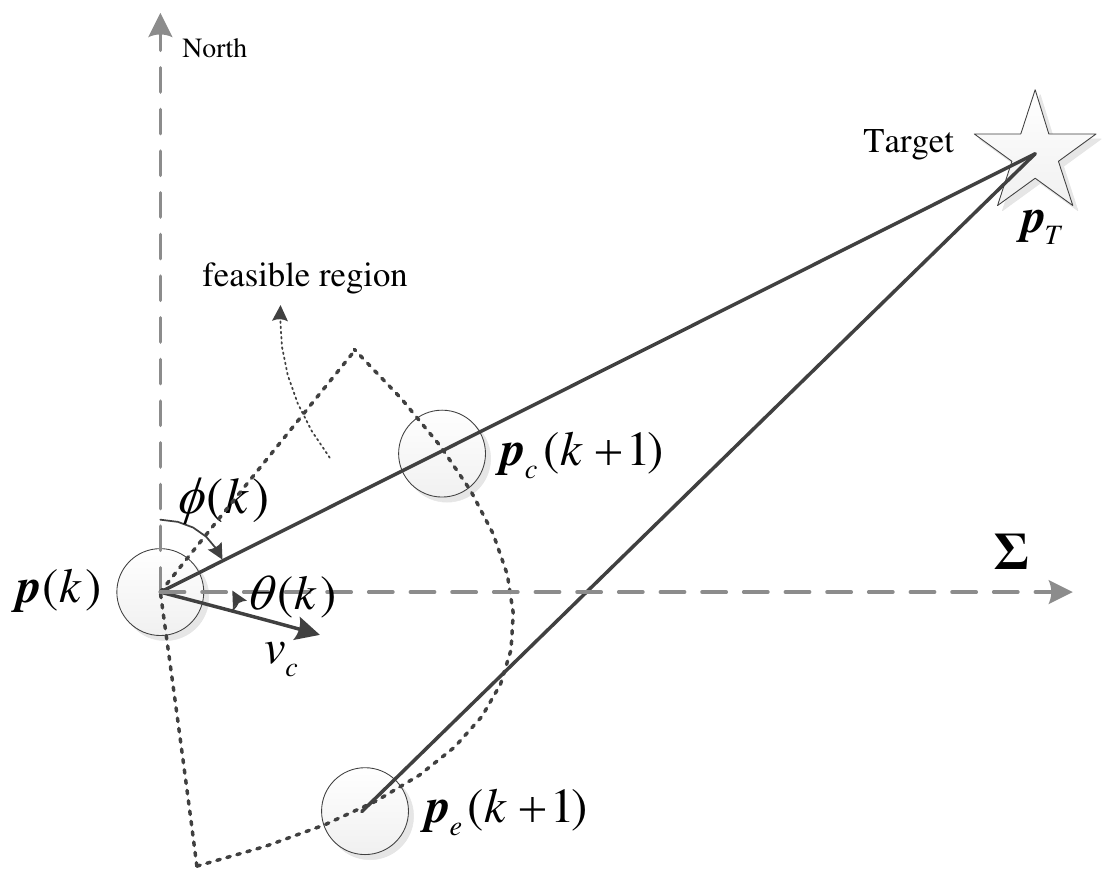} 
\caption{The bearing-only target search problem with a mobile vehicle.}
\label{fig:prob}
\end{figure}

Consider the scenario in Fig. \ref{fig:prob}, where a target is located at an unknown position $\bfp_T=[x_T, y_T]'\in\bR^2$ and the vehicle's position is $\bfp(k)=[x(k), y(k)]'\in\bR^2$ at the $k$-th time step in the global coordinate $\bf{\Sigma}$. We first assume that the GPS information $\bfp(k)$ is available and then remove it in Section \ref{sec:controller2}. The vehicle measures a noisy bearing angle $m(k)$ from itself to the target, i.e.,   
\beq \label{eqn:measure}
m(k)= \phi(k)+e(k),
\enq
where $\phi(k)=\arctan\big(({x(k)-x_T})/({y(k)-y_T})\big)$ denotes the true azimuth bearing angle, and $e(k)$ denotes the measurement noise. 

In this work, we adopt the following discrete-time Dubins vehicle \cite{liu2015discrete} 
\bee \label{eqn:model}
\begin{aligned}
x(k+1)&=x(k)+\frac{2v_c}{\omega(k)}\sin\Big(\frac{\omega(k) h}{2}\Big)\cos\Big(\theta(k)+\frac{\omega(k) h}{2}\Big),\\
y(k+1)&=y(k)+\frac{2v_c}{\omega(k)}\sin\Big(\frac{\omega(k) h}{2}\Big)\sin \Big(\theta(k)+\frac{\omega(k) h}{2}\Big),\\
\theta(k+1)&=\theta(k)+\omega(k) h
\end{aligned}
\ene
to search the target, where $v_c$ is the constant forward velocity,  $\omega(k)$ is the angular velocity to be tuned, $\theta(k)$ is the orientation, and $h$ is the sampling period. Since the forward velocity $v_c$ is finite, the vehicle can only move within a certain region in a sampling period, which is called the feasible region, see Fig. \ref{fig:prob}. The bearing-only target search problem is to sequentially design the angular velocity $\omega(k)$ only by the noisy bearing angles, such that the vehicle {\em approaches} the {\em unknown} target position as fast as possible.  

Obviously, if $\bfp_T$ were known to the vehicle, the optimal angular velocity is the one that drives the vehicle to the closest waypoint to the target within the feasible region, i.e., $\bfp_c(k+1)$, which is called the control objective. Whereas this waypoint is not achievable for $\bfp_T$ is unknown and needs to be estimated. In light of \cite{bishop2007optimality}, the optimal next waypoint to optimize the estimation performance for the unknown position $\bfp_T$ is illustrated as $\bfp_e(k+1)$ in the feasible region, which is called the estimation objective. However, this waypoint even cannot ensure the vehicle to approach the target. 

Based on these observations, the optimal angular velocity $\omega^*(k)$ should simultaneously consider both the estimation and the control objectives. To this end, we formulate the following bi-objective optimization problem
\beq  \label{eqn:opt}
\begin{aligned}
\operatorname*{\text{maximize}} \limits_{\omega(k)} &~~
\begin{bmatrix}   
    f_e(\bfp(k+1)) \\  
    f_c(\bfp(k+1)) 
  \end{bmatrix}
  \text{subject to}~\eqref{eqn:model},
  \end{aligned}
\enq
where the objective functions $f_e(\bfp(k+1))$ and $f_c(\bfp(k+1))$ represent the estimation and the control objectives, respectively. Notice that both objectives depend on the unknown position ${\bf p}_T$ and are essential to the target search problem.

\subsection{The controller architecture}

\begin{figure}
  \centering
\includegraphics[width=8.5cm]{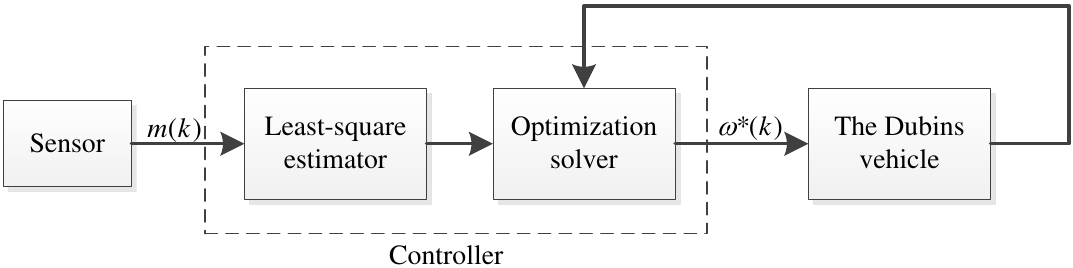} 
\caption{The controller architecture for bearing-only target search.}
\label{fig:system}
\end{figure}

To solve the bearing-only target search problem, we design a controller for the mobile vehicle including a recursive least-square estimator and an optimization solver. Fig. \ref{fig:system} shows the proposed controller architecture. Firstly, the online measurement $m(k)$ is utilized to estimate the unknown target position by the recursive least-square approach. Then, we solve the bi-objective optimization problem \eqref{eqn:opt} in ties with the estimate and obtain the angular velocity $\omega^*(k)$. 
The details of the controller design are elaborated in later sections.

\section{Solving the bi-objective optimization problem}
\label{sec:reformulation}

To solve the bi-objective optimization problem (\ref{eqn:opt}), we adopt the weighted sum method \cite{marler2010weighted} and combine the two objective functions of \eqref{eqn:opt} into the following composite form
\beq \label{eqn:com1}
f_e(\bfp(k+1))+\alpha f_c(\bfp(k+1)),
\enq
where the weighting parameter $\alpha>0$ plays an important role in balancing the two objectives, {but lacks physical meaning.}

In this section, we first describe the concrete expressions of the two objective functions $ f_e(\bfp(k+1))$ and $f_c(\bfp(k+1))$, and then solve the optimization problem.

\subsection{The estimation and the control objective functions}
Assume the measurement noise $e(k)$ in (\ref{eqn:measure}) is white Gaussian noise with zero mean and $\sigma^{2}$ variance, i.e., $e(k)\sim\cN(0,\sigma^{2})$. The estimation performance is commonly evaluated by the Cramer-Rao lower bound (CRLB) $C(\bfp_T)$ \cite{trees2003detection}. We utilize the most recent $n$ measurements at the current time step $k \geqslant n$, denoted as $\bfm_{n}(k)=[m(k-n+1), m(k-n+2), \ldots, m(k)]'$, to measure the CRLB, and we adopt the D-optimal design. In light of Theorem 1 in \cite{bishop2007optimality}, the estimation objective at the time step $k$ is explicitly expressed as
\bee \label{eqn:FIM1}
\begin{aligned}
f_e(\bfp(k+1)) =\det(C(\bfp_T)^{-1})
=\frac{1}{\sigma^4}\sum_{i<j} \frac{\sin^2(\phi(j)-\phi(i))}{r^2(i)r^2(j)},
\end{aligned}
\ene
where $ i,j \in \{k-n+1, \ldots, k+1\}$ and $r(\cdot)=\|\bfp(\cdot)-\bfp_T \|$ is the Euclidean distance from $\bfp(\cdot)$ to $\bfp_T$, i.e., the range between the vehicle and the target.

For the control performance of the vehicle approaching the target, we consider to use the range difference within a sampling period, which evaluates the approaching speed of the vehicle to the target. We express the control objective by maximizing the sum of the difference of the squared range  in $n$ sampling periods
\beq \label{eqn:per_range}
f_c(\bfp(k+1)) = \sum_{i=k-n+2}^{k+1}\left(r^2(i-1)-r^2(i)\right)
\enq
under the constraint $r(i) \leqslant r(i-1)$, since the vehicle is always required to approach the target. Observe that we employ the difference of the squared range in \eqref{eqn:per_range} instead of the absolute range difference for the ease of solving the optimization problem. 

For brevity, we only utilize the vehicle's current position and measurements to design the angular velocity, i.e., $n=1$ in \eqref{eqn:FIM1} and \eqref{eqn:per_range}. Then the composite optimization problem in \eqref{eqn:opt}-\eqref{eqn:per_range} is expressed as
\bee \label{eqn:opt_two}
\begin{aligned}
\operatorname*{\text{maximize}}\limits_{\omega(k)} &~
    \displaystyle{\frac{\sin^2 \left( \phi(k+1)-\phi(k)\right)}{\sigma^4 {r^2(k)}{r^2(k+1)}}}+\alpha \big( r^2(k)-r^2(k+1)\big)\\
\text{subject to} & ~~ r({k+1}) \leqslant r(k)~\text{and}~(\ref{eqn:model}),
\end{aligned}
\ene 
where the first constraint ensures the vehicle to gradually approach the target, and the second constraint results from the dynamics of the Dubins vehicle.

\begin{rem} Both objective functions \eqref{eqn:FIM1} and \eqref{eqn:per_range} depend on the measurements ${\bf{m}}_{n}(k)$, which are utilized to estimate the bearing angles and the ranges. In fact, a small number of the measurements, i.e., ${\bf{m}}_{n}(k)$ with a small $n$, are sufficient to evaluate both estimation and control performance at the $k$-th time step, since it has weak correlation with the measurement at the $i$-th time step for $i\ll k$. In this work, we consider a simple case of $n=1$.

\end{rem}

In the rest of this section, we show how to solve the composite optimization problem \eqref{eqn:opt_two}.

\subsection{Transformation and normalization of the composite optimization problem (\ref{eqn:opt_two})}
\label{sub:trans_nor}

\subsubsection{Transformation} 
\begin{figure}
  \centering
\includegraphics[width=8cm]{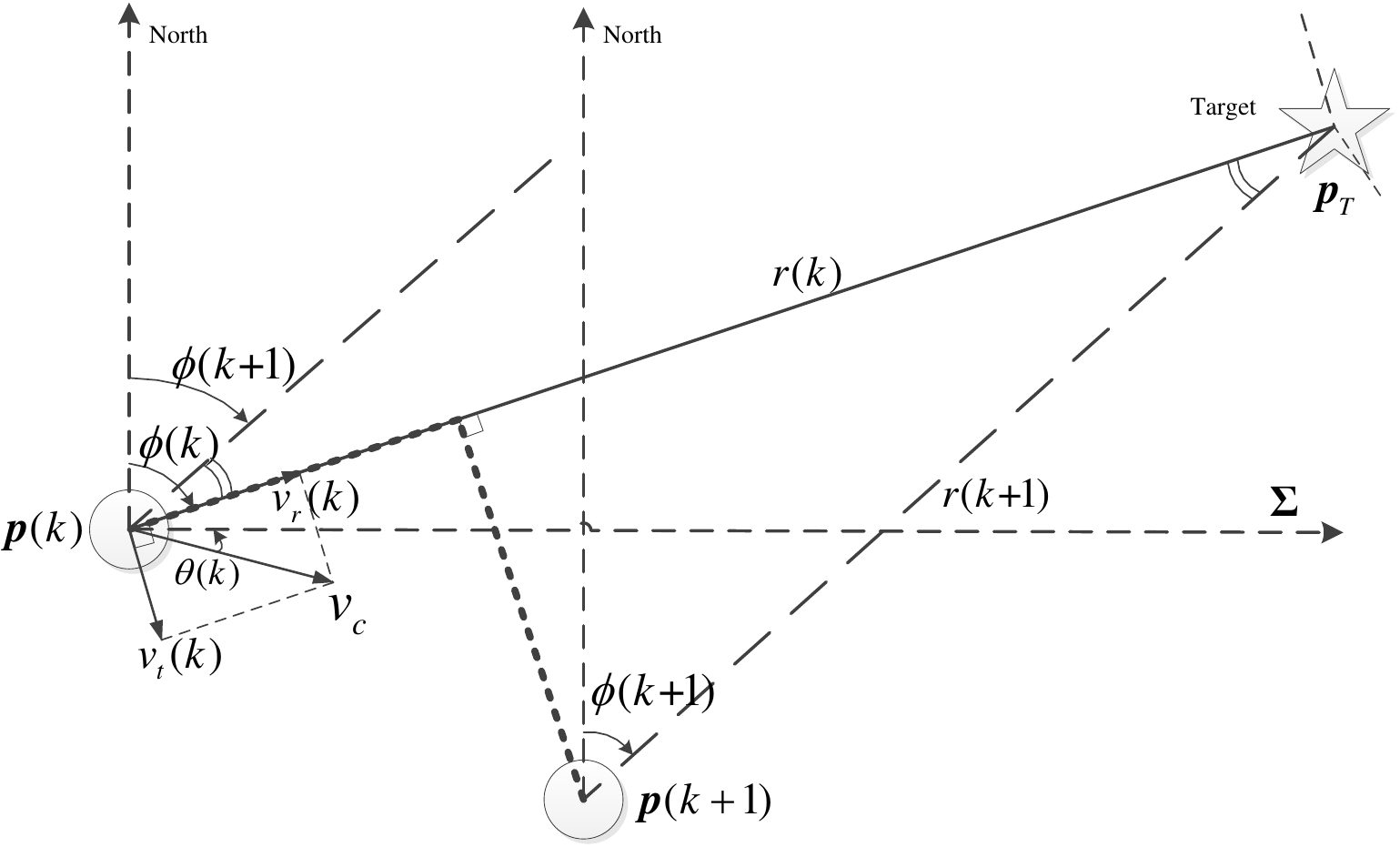} 
\caption{The decomposition of the forward velocity and geometrical relationship of two sequential waypoints.}
\label{fig:u_k1k2}
\end{figure}

Observe that it is very difficult to directly solve the composite optimization problem \eqref{eqn:opt_two}. We transform it into a solvable form by decomposing the forward velocity $v_c$ into $v_r(k)$ and $v_t(k)$, which denote the radial and tangential velocities of the vehicle relative to the target at the time step $k$, respectively, see Fig. \ref{fig:u_k1k2}.  

Then, the following relation is clear from Fig. \ref{fig:u_k1k2} that
\bee \label{eqn:v_geo}
v_c\begin{bmatrix}\cos(\theta(k))\\ \sin(\theta(k)) \end{bmatrix}  = v_r(k)  \begin{bmatrix} \sin\phi(k) \\  \cos\phi(k) \end{bmatrix} +v_t(k)   \begin{bmatrix} \cos\phi(k) \\ -\sin\phi(k) \end{bmatrix}.
\ene
Assuming that $v_r(k), v_t(k)$ and $\phi(k)$ are constant within the sampling period from $kh$ to $(k+1)h$, we take Euler integral with respect to the continuous dynamics on the right side of \eqref{eqn:v_geo} and obtain that
\bee \nonumber
\begin{aligned}
x(k+1)-x(k)&=h\big(v_r(k) \sin(\phi(k))+v_t(k) \cos(\phi(k))\big),\\
y(k+1)-y(k)&=h\big(v_r(k) \cos(\phi(k))-v_t(k) \sin(\phi(k))\big).
\end{aligned}
\ene  
Together with the Dubins vehicle dynamics \eqref{eqn:model}, it holds that
\beq
\begin{aligned} \label{eqn:omega}
\frac{2v_c}{\omega(k)}\sin\Big(\frac{\omega(k) h}{2}\Big)&\cos\Big(\theta(k)+\frac{\omega(k) h}{2}\Big)\\
= &h\big(v_r(k) \sin(\phi(k))+v_t(k) \cos(\phi(k))\big),\\
\frac{2v_c}{\omega(k)} \sin\Big(\frac{\omega(k) h}{2}\Big) & \sin\Big (\theta(k)+\frac{\omega(k) h}{2}\Big)\\
= &h \big( v_r(k) \cos(\phi(k))-v_t(k) \sin(\phi(k)) \big).
\end{aligned}
\enq
Furthermore, dividing the two equalities in \eqref{eqn:omega} leads to that
\bee \nonumber
\tan \big(\theta(k)+\frac{\omega(k) h}{2}\big)=\frac{v_r(k) \cos(\phi(k))-v_t(k) \sin(\phi(k))}{v_r(k) \sin(\phi(k))+v_t(k) \cos(\phi(k))},
\ene
which implies the relation between $\omega(k)$ and $v_r(k), v_t(k)$, i.e.
\beq \label{eqn:omega_tan}
\begin{aligned}
\omega(k)&= \frac{2}{h} \arctan \Big(  \frac{v_r(k) \cos(\phi(k))-v_t(k) \sin(\phi(k))}{v_r(k) \sin(\phi(k))+v_t(k) \cos(\phi(k))}  \Big)\\
&~~-\frac{2}{h}\theta(k).
\end{aligned}
\enq

Moreover, the decomposition of $v_c$ shows that $v_r(k)h$ is exactly the projected distance of $\bfp(k+1)-\bfp(k)$ onto the direction of $\bfp_T-\bfp(k)$, i.e.,
\bee \label{eqn:vrh}
v_r(k)h=\Big \|\frac{(\bfp_T-\bfp(k))(\bfp_T-\bfp(k))'}{\| \bfp_T-\bfp(k) \|^2}(\bfp(k+1)-\bfp(k))\Big \|,
\ene
which is denoted as the bold dashed segment on the direction of $\bfp_T-\bfp(k)$ in Fig. \ref{fig:u_k1k2}.
Similarly, $v_t(k)h$ can be represented as
\bee \label{eqn:vth}
v_t(k)h=\Big \|\Big( \text{I}_2-\frac{(\bfp_T-\bfp(k))(\bfp_T-\bfp(k))'}{\| \bfp_T-\bfp(k) \|^2}\Big)(\bfp(k+1)-\bfp(k))\Big \|,
\ene
which is denoted as the bold dashed segment on the direction perpendicular to $\bfp_T-\bfp(k)$ in Fig. \ref{fig:u_k1k2}. Jointly with \eqref{eqn:vrh}, \eqref{eqn:vth} and the triangular geometric relationship in Fig. \ref{fig:u_k1k2}, it follows that
\beq \label{eqn:geo}
\begin{aligned}
&\sin^2(\phi(k+1)-\phi(k))= \frac{v_t^2(k)}{(r(k)/h-v_r(k))^2+v_t^2(k)},\\
&r^2(k)-r^2(k+1)= 2r(k)v_r(k)h-(v_r^2(k)+v_t^2(k))h^2.
\end{aligned}
\enq
Besides, it follows from \eqref{eqn:v_geo} that
\begin{align}
\label{eqn:k1k2} 
v_r^2(k)+v_t^2(k)&=v_c^2. 
\end{align}
Substituting (\ref{eqn:geo}) and (\ref{eqn:k1k2}) into the composite optimization problem (\ref{eqn:opt_two}) yields that
\bee \label{eqn:opt2}
\begin{aligned}
\operatorname*{\text{maximize}}\limits_{v_r} & ~~ f_e(v_r)+\alpha f_c(v_r)\\
\text{subject to} & \ \frac{v_c^2h}{2r} \leqslant v_r \leqslant v_c,
\end{aligned}
\ene
where the dependence of $v_r(k)$ and $r(k)$ on $k$ is dropped for simplifying notations and  
\beq \label{obj_ec}
\begin{aligned}
f_e(v_r)&= \frac{(v_c^2-v_r^2)h^2}{\sigma^4r^2(r^2-2rv_rh+v_c^2h^2)^2},\\
f_c(v_r)&= 2rv_rh-v_c^2h^2.
\end{aligned}
\enq

\subsubsection{Normalization}  Observe that $f_e(v_r)$ and $f_c(v_r)$ are used to quantify the objectives with different physical meanings, which results in the lack of physical understanding of the weighting parameter $\alpha$. Consequently, it is unclear how to provide an appropriate $\alpha$ to balance the estimation and the control objectives. To overcome this difficulty, we normalize the two objective functions in \eqref{obj_ec} to be dimensionless and with values in $[0,1]$. Then, the composite optimization problem \eqref{eqn:opt2} is rewritten in the following normalized form
\beq \label{eqn:opt3}
\begin{aligned}
\operatorname*{\text{minimize}}\limits_{v_r} \ f(v_r)~
\text{subject to} & \ \frac{v_c^2h}{2r} \leqslant v_r \leqslant v_c,
\end{aligned}
\enq
where the normalized objective function
\begin{eqnarray} 
f(v_r)& =& \Big(1-\frac{(v_c^2-v_r^2)(r^2-v_c^2h^2)^2}{v_c^2(r^2-2rv_r h+v_c^2h^2)^2}\Big)\nonumber\\
&&+\beta \ \frac{ (v_c-v_r)(v_c^2h^2+r^2)}{v_c(v_ch-r)^2}\label{objopt3}
\end{eqnarray}
introduces a new weighting parameter $\beta \in [0, +\infty)$. See Appendix \ref{app:A} for the detailed normalization process.

In contrast with the weighting parameter $\alpha$ in \eqref{eqn:opt2}, the new weighting parameter $\beta$ in \eqref{objopt3} can directly quantify the relative importance of the control objective to the estimation objective, which is also an advantage of the normalized optimization problem \eqref{eqn:opt3} over those in \cite{ergezer2013path, duan2014pigeon, sidoti2017a, maurovi2018path} with simple weighted sums as the objective functions. To be more specific, $\beta=1$ in \eqref{objopt3} means the control objective has the same importance as the estimation objective, while the physical meaning of $\alpha=1$ is unclear. Therefore, the new weighting parameter $\beta$ facilitates designers to emphasize the importance between the estimation and the control objectives.

Furthermore, solutions to the normalized optimization problems with different  values of $\beta$  lead the vehicle to move in different trajectories. It is interesting to note that if $\beta\geqslant 4$, the solution of \eqref{eqn:opt3} results in the heading direction of the vehicle directly toward the estimate of the target position. In this case, the controller design ignores the estimation objective as in \cite{deghat2014localization, guler2017adaptive, liu2018a}.  

\begin{rem}
We set a constant $\beta$ in this work. In fact, it can be seen as an optimization variable to be adjusted during the target search task, which, however, is beyond of the scope of this work. A heuristic idea is to design $\beta(k) \propto 1/r(k)$, since the target position may have already been well estimated when the vehicle is close to the target. Another idea is to measure estimation progress by using the following ratio
$$
\frac{f_e(v_r(k)))}{f_e(v_r(k-1))}
$$
and increase $\beta(k)$ if this ratio is close to unity, and decrease it otherwise. 

\end{rem}

\subsection{Solution to the normalized optimization problem \eqref{eqn:opt3}}

Before providing the solution, we introduce an auxiliary variable $$\rho=v_ch/r$$ to denote the level of completing the target search task. Specifically, $\rho>1$, i.e., $v_ch > r$, means that the vehicle's traveling distance in a sampling period is strictly larger than its distance to the target. This implies that the vehicle is sufficiently close to the target. Hence, we only focus on the nontrivial case that $0<\rho\leqslant1$. 

\begin{lemma} \label{lemma}
The equation  $\frac{\text{d}}{\text{d}v_r}{f}(v_r)=0$ has at least one root in the interval $[v_s,v_c]$, where $v_s={2rv_c^2h}/({r^2+v_c^2h^2})$.  
\end{lemma}

\begin{prf}
See Appendix \ref{app:B}. \hfill $\blacksquare$
\end{prf}

Let $\rho_c=({2-\sqrt{4\beta-\beta^2}})/({\beta-2})$ and $v_z$ denote the largest root of the equation $\frac{\text{d}}{\text{d}v_r}{f}(v_r)=0$, $v_r \in [v_s,v_c] $ by Lemma \ref{lemma}. Then, we provide the solution of \eqref{eqn:opt3} in the following theorem.

\begin{thm} \label{thm:1}

The optimal solution $v_r^*$ of the normalized optimization problem \eqref{eqn:opt3} is given as follows.

\begin{enumerate}
\renewcommand{\labelenumi}{\rm(\alph{enumi})}
\item 
If $\beta \in [4, \infty)$, then $v_r^*=v_c$.
\item
If $\beta \in [2,4)$, then  
$$v_r^*=\left \{
\begin{array}{ll}
v_c,&\text{when}~\rho \in (0,\rho_c],\\
v_z,&\text{when}~\rho \in (\rho_c, 1].
\end{array}
\right.
$$
\item
If $\beta \in [1,2)$, then $v_r^*=v_z$.
\item
If $\beta \in [0,1)$, then
$$v_r^*=\left \{
\begin{array}{ll}
v_z,&\text{when}~f(v_z)<\beta,\\
v_s,&\text{otherwise}.
\end{array}
\right.
$$

\end{enumerate}
\end{thm}

\begin{prf}
See Appendix \ref{app:C}. \hfill $\blacksquare$
\end{prf}
Note that $v_s \leqslant v_z\leqslant v_c$. Theorem \ref{thm:1} shows that a larger $\beta$ results in a larger $v_r^*$. As we discussed in Section \ref{sub:trans_nor} that a larger $\beta$ emphasizes more the importance of the control objective, a larger radial velocity $v_r$ is derived to approach the target faster. This is also consistent with our intuition. Specifically, if $\beta \in [4,\infty)$, then $v_r^*$ directly takes its maximum value $v_c$, which implies that the vehicle almost ignores the estimation objective.
 
In addition, observe that finding $v_z$ is not needed in all cases in Theorem \ref{thm:1}. This potentially saves computational cost, compared to  the works in \cite{ergezer2013path, duan2014pigeon, sidoti2017a, maurovi2018path, Candido2011Minimum}.

\section{The optimization-based controller }
\label{sec:controller}

It should be noted from Theorem \ref{thm:1} and \eqref{eqn:omega_tan} that the vehicle's optimal angular velocity depends on the range $r(k)$ and the bearing angle $\phi(k)$, both of which are unknown and need to be estimated. Accordingly, we design a recursive least-square estimator, and utilize the certainty-equivalence principle for the controller design.

\subsection{The recursive least-square estimator}

Observing that the unknown $r(k)$ and $\phi(k)$ are both originated from the unknown position $\bfp_T$, we only need to obtain an optimal estimate of $\bfp_T$, which is the solution of the following nonlinear least-square minimization problem
\beq \label{eqn:est}
\operatorname*{\text{minimize}}\limits_{ \bfp_T}~ \frac{1}{2} \sum_{i=1}^k \left\| m(i)-\text{arctan}\Big(\frac{x(i)- x_T}{y(i)- y_T}\Big) \right\|^2.
\enq
Similar to Stansfield estimator in\cite{gavish1992performance}, we also assume the measurement noise $e(k)$ is sufficiently small to justify the replacement of $m(i)-\arctan (({x(i)- x_T})/({y(i)- y_T)} )$ with 
$\sin \big( m(i)-\arctan ((x(i)- x_T)/(y(i)- y_T) )\big).$ 
Using the relation that
\beq \label{eqn:app_sin}
\begin{aligned}
&\sin \Big( m(i)-\arctan \big(\frac{x(i)- x_T}{y(i)- y_T}  \big)\Big)
\\&=\frac{1}{r(i)}\big( \sin(m(i))(y(i)-y_T) - \cos(m(i))(x(i)-x_T) \big),
\end{aligned}
\enq 
we approximate the optimization problem \eqref{eqn:est} as 
\beq \label{eqn:app_opt}
\begin{aligned}
\operatorname*{\text{minimize}}\limits_{\bfp_T}~ \frac{1}{2} \sum_{i=1}^k &\| x(i)\cos (m(i))-y(i) (\sin (m(i)) \\&- [\cos (m(i)) , -\sin(m(i))] [x_T, y_T]' \|^2,
\end{aligned}
\enq
where we ignore the rather weak effect of $1/r(i)$ in \eqref{eqn:app_sin}. 

Since the optimization problem \eqref{eqn:app_opt} is a standard least-square problem, we design a recursive algorithm to solve it, see Algorithm \ref{alg:estimate}. In practice, the initial estimate of the target position $\widehat \bfp_{T}(1)$ in Algorithm \ref{alg:estimate} is given as a solution of the following equations
\bee\label{initial}
\tan (m(i))=\frac{x(i)- x_T}{y(i)- y_T}, i=0,1. 
\ene
\begin{algorithm}
\caption{Least-square estimate}
\begin{algorithmic} [1]
\STATE At time $k=1$, set $\widehat \bfp_T(1)$ as the solution of \eqref{initial}, and $\bfQ(1)=\text{I}_2.$
\STATE At time $k\geqslant 2$, the vehicle takes a bearing angle $m(k)$ and computes
$$
\begin{aligned}
\bfH&=[-\cos(m(k)), \sin(m(k))]',\\
\bfK&=\bfQ(k-1) \bfH( \bfH'\bfQ(k-1) \bfH+1)^{-1},\\
e_T&= \bfH'\bfp(k),\\
\bfQ(k) &= (\text{I}_2- \bfK \bfH')\bfQ(k-1),\\
\widehat \bfp_T(k) &= \widehat \bfp_T(k-1) + \bfK(e_T- \bfH' \widehat \bfp_T(k-1)).
\end{aligned}
$$
\end{algorithmic}
\label{alg:estimate}
\end{algorithm}

Note that Algorithm \ref{alg:estimate} is developed under the assumption of small $e(k)$ and ignorance of $1/r(k)$, and might be biased. Nonetheless, it is verified by the simulation results in \cite{douganccay2004bias} to perform well with a small estimation error and to be of practical significance.

\subsection{The controller for bearing-only target search}
Now, the detailed controller providing the angular velocity is shown in Algorithm \ref{alg:control}.

\begin{algorithm}
\caption{The angular velocity for the vehicle}
\begin{algorithmic} [1]
\STATE  At time $k = 0$, the vehicle takes a bearing angle $m(k)$, and randomly selects $\omega(k)$.
\STATE At time $k\geqslant 1$, the vehicle takes a bearing angle $m(k)$ and completes the following steps:
\begin{enumerate}
\renewcommand{\labelenumi}{\rm(\alph{enumi})}
\item runs Algorithm \ref{alg:estimate} to obtain $$\widehat \bfp_{T}(k)=[\widehat x_T(k), \widehat y_T(k)]',$$
\item computes both the bearing and range estimates
$\begin{aligned}\widehat \phi(k)&= \arctan\big((x(k)-\widehat x_T(k))/(y(k)-\widehat y_T(k))\big),\\ 
\widehat r(k)&= \|\bfp(k)-\widehat \bfp_T(k) \|,
\end{aligned}$
\item uses Theorem \ref{thm:1} to compute $v_r^*(k)$ where $r(k)$ is replaced by its estimate $\widehat r(k)$,
\item computes the angular velocity in \eqref{eqn:omega_tan} by replacing $\phi(k)$ and $v_r(k)$ by $\widehat\phi(k)$ and $v_r^*(k)$, respectively. 
\end{enumerate}

\end{algorithmic}
\label{alg:control}
\end{algorithm}

\section{{Control for target search without GPS information}}
\label{sec:controller2}

{Observe that the controller in Algorithm \ref{alg:control} requires the vehicle's position $\bfp(k)$ in the global coordinate, which is unavailable in GPS-denied environments. In this section, we consider the target search problem without the vehicle's global position, and utilize the local bearing angles defined in the vehicle's coordinate to solve it.}

\subsection{ {The target search problem without GPS information}}
\begin{figure}
  \centering
\includegraphics[width=7cm]{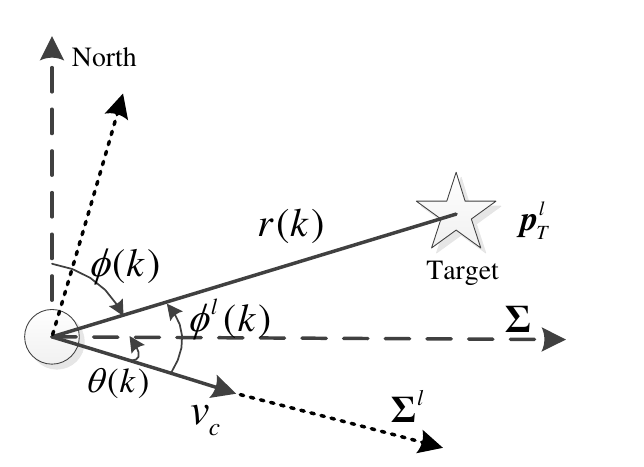} 
\caption{The bearing-only target search problem in the vehicle's coordinate.}
\label{fig:problem2}
\end{figure}

Without the vehicle's global position $\bfp(k)$, we consider the target search problem in the vehicle's coordinate ${\bf{\Sigma}}^l$. As Fig. \ref{fig:problem2} shows, this coordinate is attached to the vehicle with its positive x-axis coincident with the heading of the vehicle. The target's position in this coordinate is denoted as $\bfp_T^l(k)=[x_T^l(k), y_T^l(k)]'$ at the time step $k$, which is time-varying due to the moving vehicle's coordinate. The orientation $\theta(k)$ is still accessible by a compass.
Moreover, the vehicle measures a local bearing angle $m^l(k)$ in the vehicle's coordinate relative to the target, i.e.,
\beq \label{eqn:mea2}
m^l(k)= \phi^l(k)+e^l(k),
\enq
where $\phi^l(k)=\arctan(y_T^l(k)/x_T^l(k))$ denotes the true local bearing angle, and $e^l(k)\sim\cN(0,\sigma^{2})$ denotes the local measurement noise.

Then, the bearing-only target search problem becomes to sequentially design the angular velocity $\omega(k)$ for the vehicle \eqref{eqn:model} to approach the unknwon target position as fast as possible only by local noisy bearing angles. To solve this problem, we still design the controller depending on the solution to the bi-objective optimization problem \eqref{eqn:opt}. 

\subsection{{The optimization-based controller without GPS information}}

Similar to Section \ref{sec:controller}, a recursive least-square algorithm is designed in the proposed controller to estimate the target position $\bfp^l_T(k)$. We approximately solve the nonlinear least-square optimization problem
$$
\operatorname*{\text{minimize}}\limits_{\bfp_T^l(k)}~ \frac{1}{2} \sum_{i=1}^k \| m^l(i)-\arctan\Big(\frac{ x_T^l(i)}{ y_T^l(i)}\Big) \|^2
$$
by a linear least-square form, i.e., 
\bee  \label{eqn:ls_local}
 \operatorname*{\text{minimize}}\limits_{ \bfp_T^l(k)}~ \frac{1}{2} \sum_{i=1}^k \| [\cos(m^l(i)) , -\sin(m^l(i))] [x_T^l(i), y_T^l(i)]' \|^2.
\ene

The difference from Section \ref{sec:controller} lies in the time-varying target position in the vehicle's coordinate due to the moving vehicle. Thus, the state equation of the target position $\bfp_T^l(k)$ is to be derived. According to the relation between global coordinate and the vehicle's coordinate, it holds that 
\bee
\bfp_T^l(k)=
\begin{bmatrix}
\begin{aligned}   
    &\cos(\theta(k)) \  &\sin(\theta(k)) \\  
    -&\sin(\theta(k))\  &\cos(\theta(k))
\end{aligned}
\end{bmatrix}(\bfp_T-\bfp(k)).
\ene
Combining with the vehicle dynamics, we further obtain that
\bee\label{sysdyn}
\bfp_T^l(k)=\bfA(k-1)\bfp_T^l(k-1)+{\bf b},
\ene
where $$\bfA(k-1)=\begin{bmatrix}
\begin{aligned}   
    1  ~~~ \omega(k-1) \\  
    -\omega(k-1) ~~~  1
\end{aligned}
\end{bmatrix}, {\bf b}=[-v_ch, 0]'.$$

Jointly with (\ref{eqn:ls_local}) and (\ref{sysdyn}), the recursive least-square estimator with only local bearing angles is given in Algorithm \ref{alg:estimate2}, where the initial estimate of the target position in the vehicle's coordinate $\widehat \bfp_{T}^l(1)$ is selected as a solution of the following equations
\bee\label{initial_local}
\tan (m^l(i))=\frac{x_T^l(i)}{y_T^l(i)}, i=0,1. 
\ene

\begin{algorithm}
\caption{Least-square estimation algorithm with local bearing angles}
\begin{algorithmic} [1]
\STATE At time $k=1$, set $\widehat \bfp_T^l(1)$ as the solution of \eqref{initial_local}, and $\bfQ(1)=\text{I}_2.$
\STATE At time $k\geqslant 2$, the vehicle takes a local bearing angle $m^l(k)$ and computes
$$
\begin{aligned}
\bfH&=[-\cos(m^l(k)), \sin(m^l(k))]',\\
\bfK&=\bfQ(k-1) \bfH( \bfH'\bfQ(k-1) \bfH+1)^{-1},\\
\bfQ(k) &= (\text{I}_2- \bfK \bfH')\bfQ(k-1),\\
\widehat \bfp_T^l(k) &= (\bfA(k-1)- \bfK \bfH') \widehat \bfp_T^l(k-1) + \bfb.
\end{aligned}
$$
\end{algorithmic}
\label{alg:estimate2}
\end{algorithm}

The angular velocity for the vehicle is also given in Algorithm \ref{alg:control} where we use Algorithm \ref{alg:estimate2} to replace Algorithm \ref{alg:estimate} and the bearing angle and the range estimates are computed as 
$$
\begin{aligned}
\widehat \phi(k)&= \frac{\pi}{2}-\theta(k)-\text{arctan}\left(\frac{\widehat x_T^l(k)}{\widehat y_T^l(k)}\right), \\
\widehat r(k)&=\| \widehat \bfp_T^l(k) \|,
\end{aligned}
$$
and the rest of Algorithm \ref{alg:control} remain unchanged. Thus, we obtain the controller for the target search problem without GPS information.

\section{Simulations}
\label{sec:simulations}

In this section, we illustrate the effectiveness of the proposed optimization-based controllers for the bearing-only target search problem. The initial state of the vehicle is set as $\bfp(0)=[0, 0]'$, $\theta(0)=0$, the constant forward velocity $v_c=4 \text{m/s}$, the target position $\bfp_T=[100, 100]'\text{m}$, and its initial estimate $\widehat \bfp_T(0)=[40, 80]'\text{m}$. We take the weighting parameter $\beta=1$ unless particular specification, which means the control objective has the same importance as the estimation objective. Moreover, the sampling period is set as $h=0.25 \text{s}$.

\subsection{Simulation results with GPS information}
\label{subsec:static}

\begin{figure}
  \centering
\includegraphics[width=8cm]{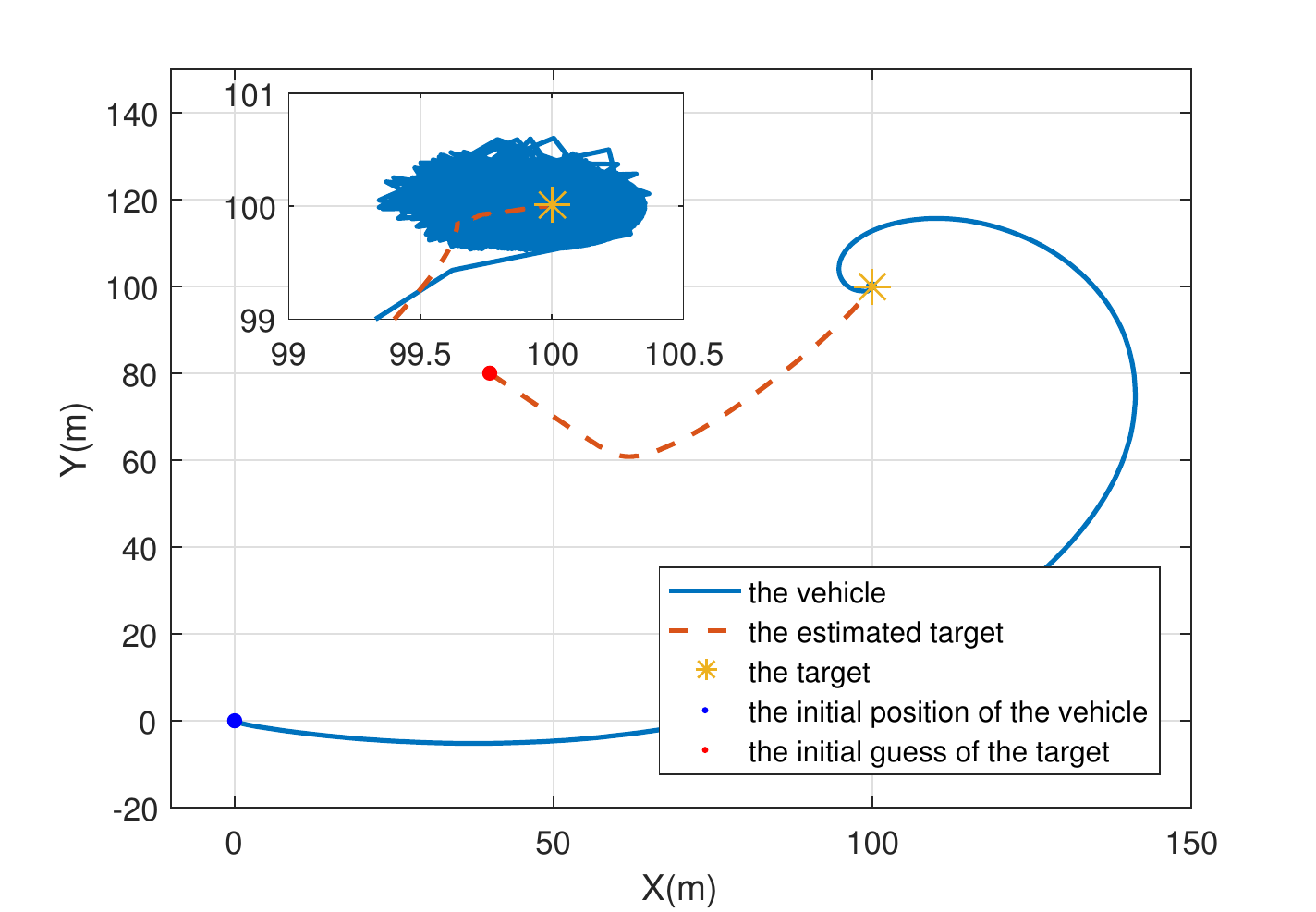} 
\caption{The trajectory of the vehicle and the estimated positions of the target. The error between the vehicle and the target is around 0.41m.}
\label{fig:tra}
\end{figure}

With GPS information, the controller in Algorithm \ref{alg:estimate} is implemented in a target search task to approach an unknown target position. Fig. \ref{fig:tra} shows the vehicle's trajectory and the estimated target positions, where the small square in the upper left corner is the magnification of the results in the target's neighbor. Clearly, the vehicle moves in a spiral curve and approaches the true target gradually, while the estimated target positions converges to its true position. It reveals that the vehicle achieves the estimation and the control objectives simultaneously with the weighting parameter $\beta=1$.

\begin{figure}
  \centering
\includegraphics[width=8cm]{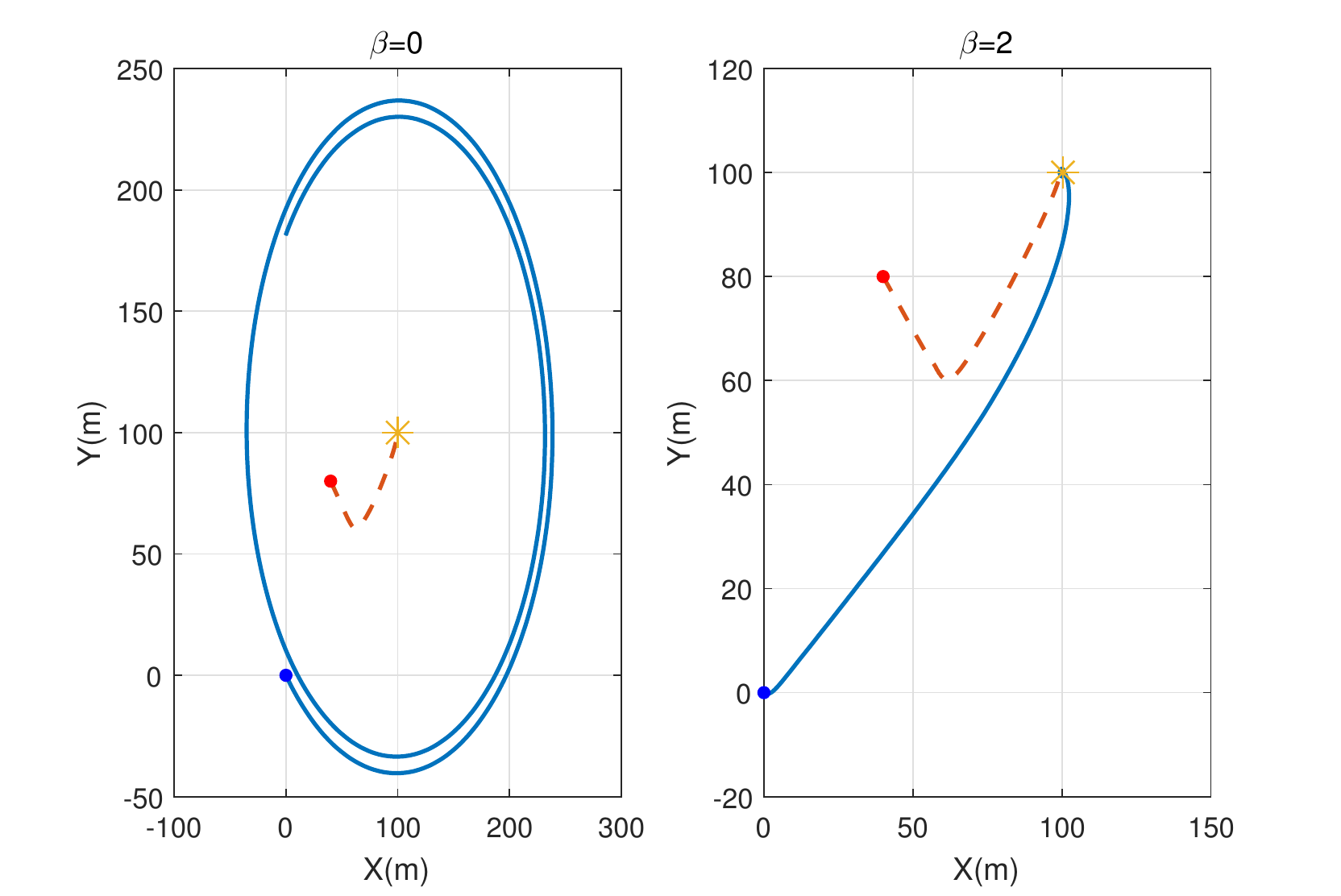} 
\caption{The trajectories of the vehicle and the estimated positions of the target with $\beta=0$ and $\beta=2$.}
\label{fig:tra_0-2}
\end{figure}

The cases of $\beta=0$ and $\beta=2$ are depicted in Fig. \ref{fig:tra_0-2}, which shows that the vehicle tends to circle the target for the estimation objective and to directly approach the target for the control objective, respectively. This also explains the rationality of the spiral trajectory in  Fig. \ref{fig:tra}, as $\beta=1$ is between these two cases. The videos of the vehicle's motions under the proposed controller in the cases of $\beta=0$, $\beta=1$, and $\beta=2$ are uploaded to \url{https://www.dropbox.com/s/9dtpijap65gup50/video_motion.mp4?dl=0.}

\subsection{Simulation results without GPS information}

\begin{figure}
  \centering
\includegraphics[width=8cm]{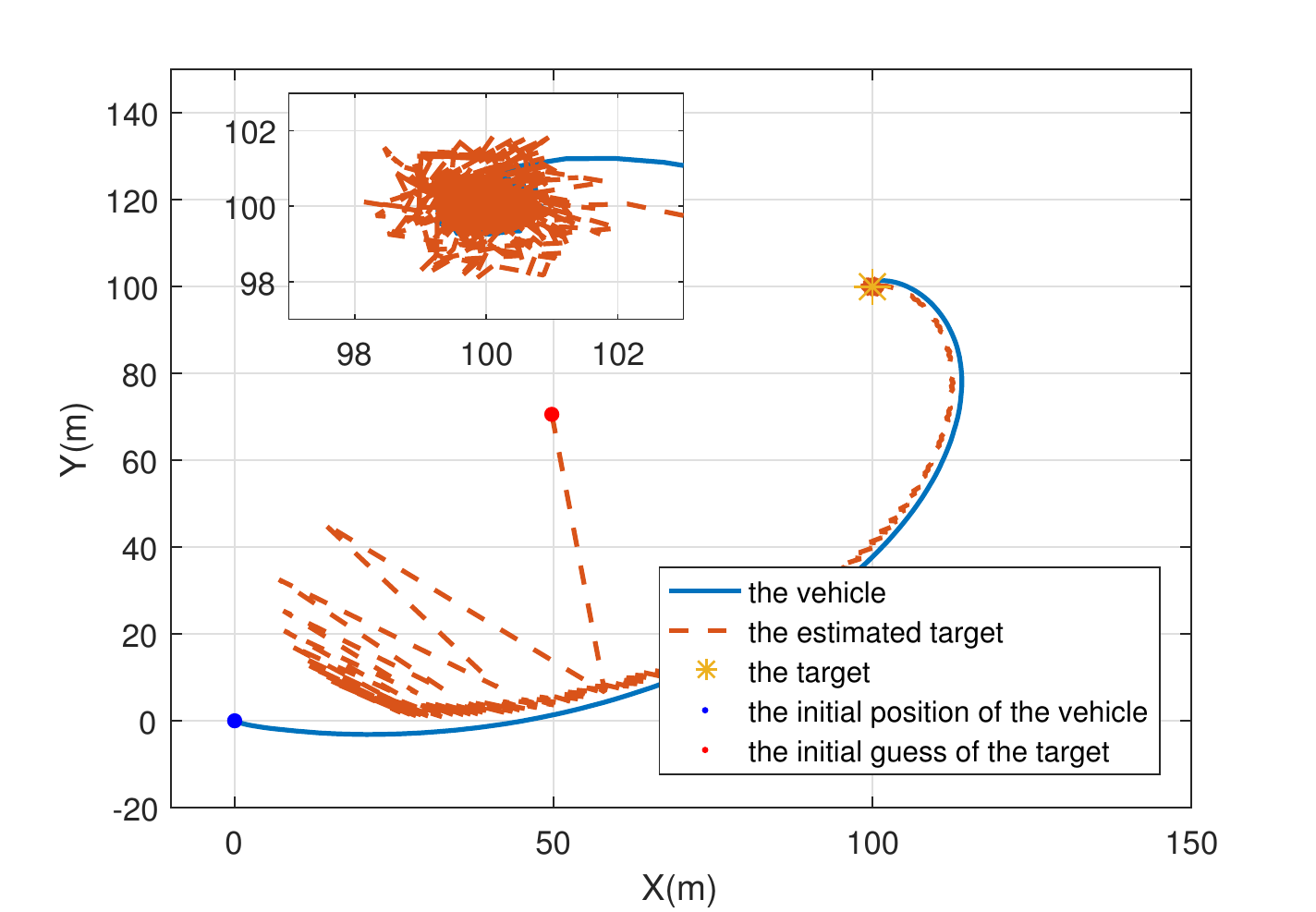} 
\caption{The trajectory of the vehicle and the estimated positions of the target with local bearing angles. The final error between the vehicle and the target is around 0.43m.}
\label{fig:tra_local}
\end{figure}

For the target search problem without the vehicle's global position $\bfp(k)$, we apply the controller in Algorithm \ref{alg:estimate2} with local bearing angles. The results are shown in Fig. \ref{fig:tra_local}, where both of the vehicle's trajectory and the estimated target positions gradually converge to the true target position. It is consistent with our results in Section \ref{sec:controller2}. Since the true target position is time-varying in the vehicle's coordinate, the estimated target positions have larger fluctuation than those in Section \ref{subsec:static}.

In addition, we compare the two recursive least-square estimators of Algorithm \ref{alg:estimate} and \ref{alg:estimate2} by denoting the estimation error as $ e_{\text{est}}(k)=\|\bfp_T(k)-\widehat \bfp_T(k) \|.$ We observe that the $e_{\text{est}}(k)$ of Algorithm \ref{alg:estimate2} converges to zero as that of Algorithm \ref{alg:estimate}, although its value is larger in the initial stage and the convergence rate is slower due to lack of the vehicle's GPS information. It implies that the GPS information is not necessary, but might improve the search performance, which is not studied in previous works \cite{bishop2007optimality, mavrommati2018real, moreno2013optimal, dogancay2012uav, meng2016communication}.

\begin{figure}
  \centering
\includegraphics[width=8cm]{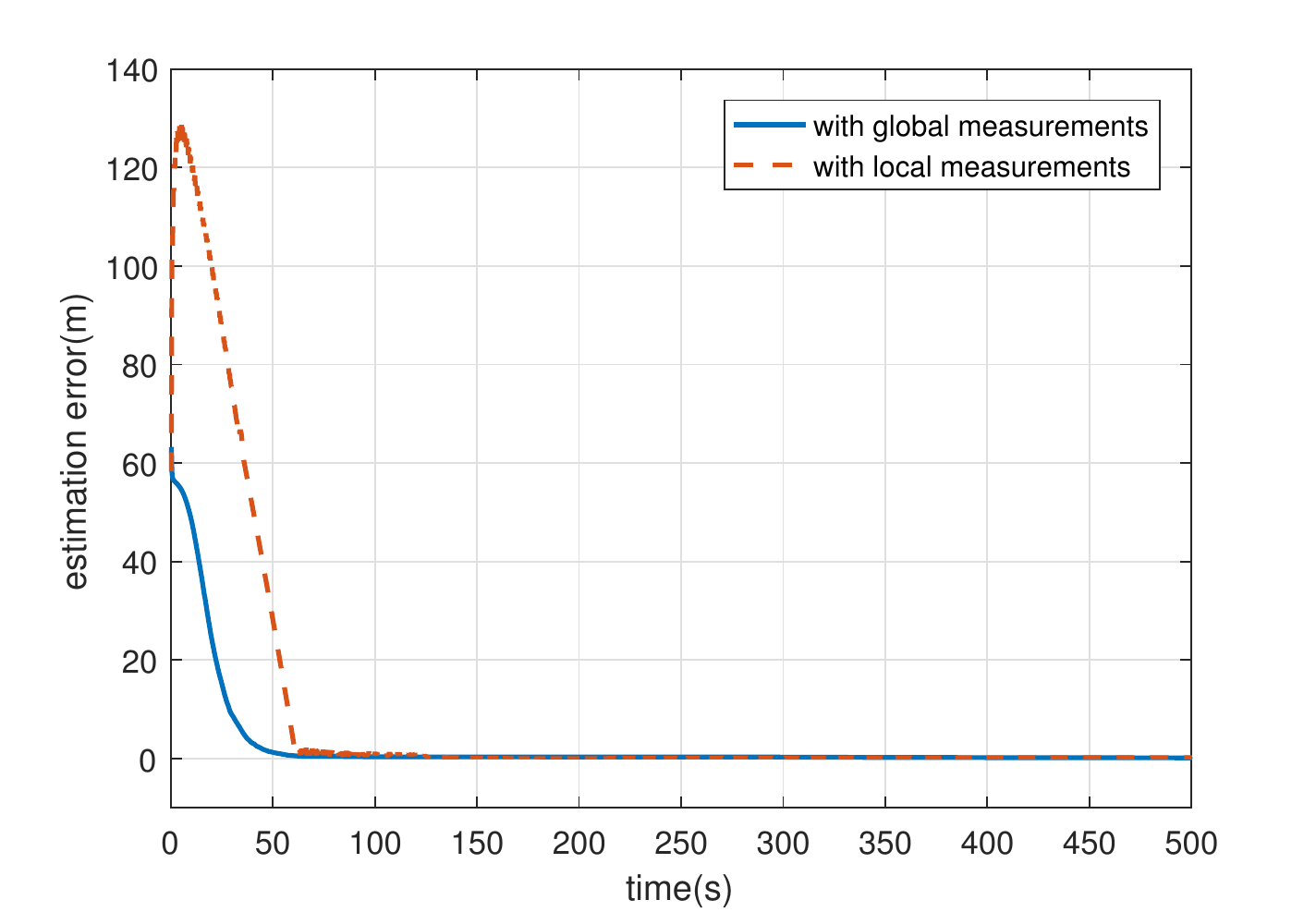} 
\caption{The estimation errors of the target position with global and local bearing angles.}  
\label{fig:mse_com}
\end{figure}

\subsection{The role of the weighting parameter $\beta$}

In this subsection, we investigate the role of the weighting parameter $\beta$ in balancing the estimation and the control objectives, and provide some guidelines for choosing an appropriate $\beta$. 

\begin{figure}
  \centering
\includegraphics[width=8cm]{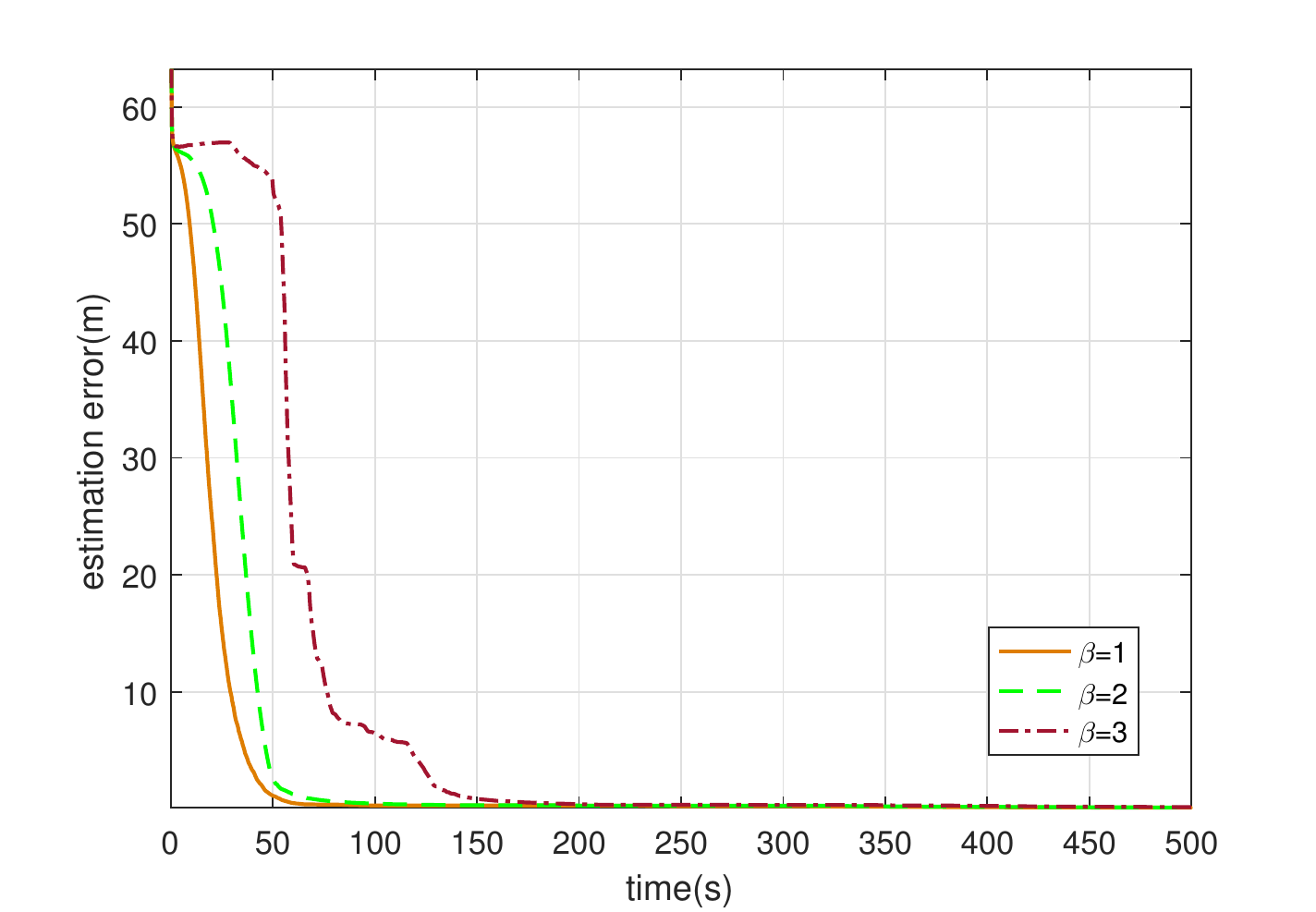} 
\caption{The estimation errors of the target position with different values of $\beta$.} 
\label{fig:e_est}
\end{figure}
\begin{figure}
  \centering
\includegraphics[width=8cm]{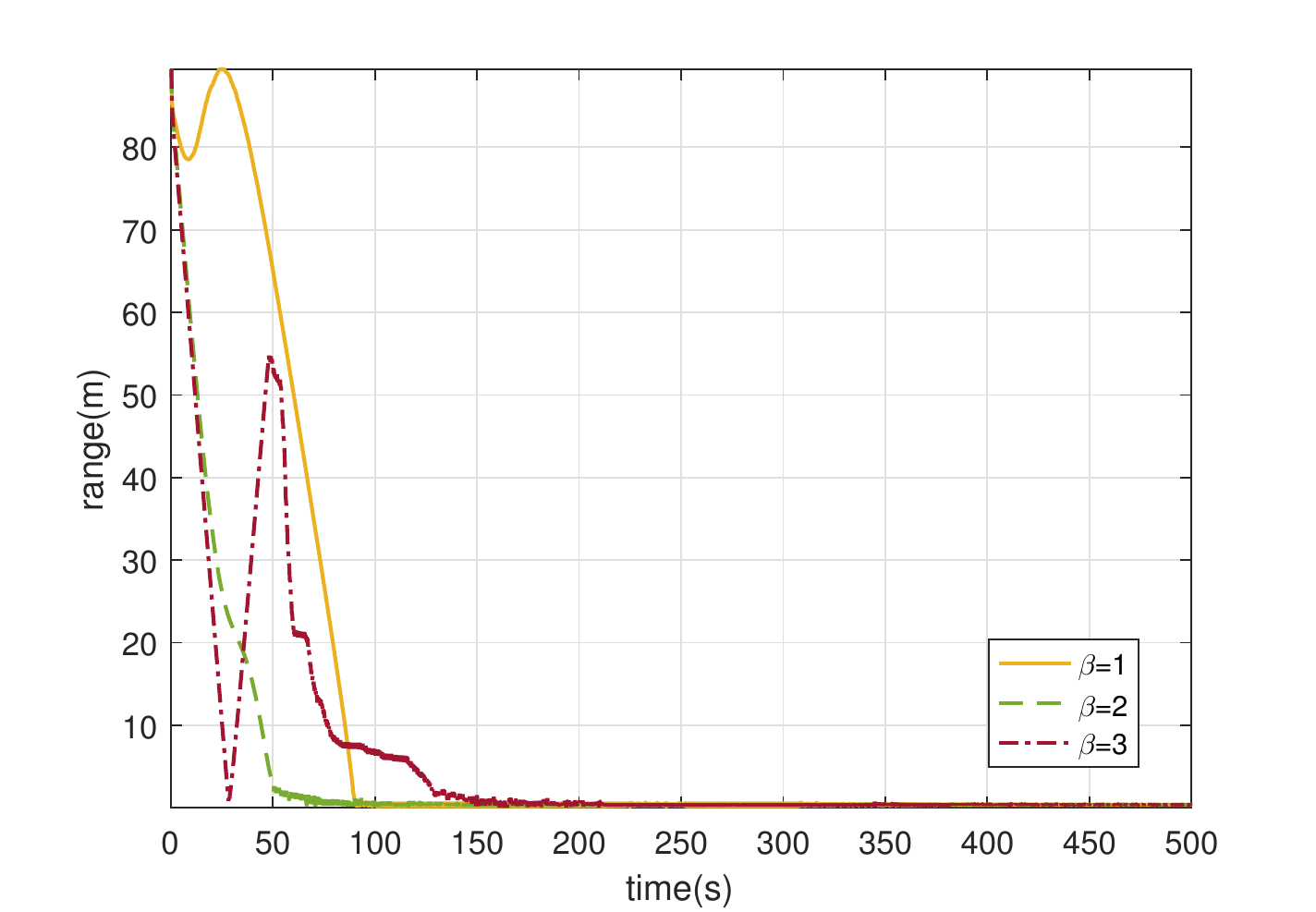} 
\caption{The ranges between the vehicle and the estimated target with different values of $\beta$.}
\label{fig:range}
\end{figure}

\begin{figure}
  \centering
\includegraphics[width=8cm]{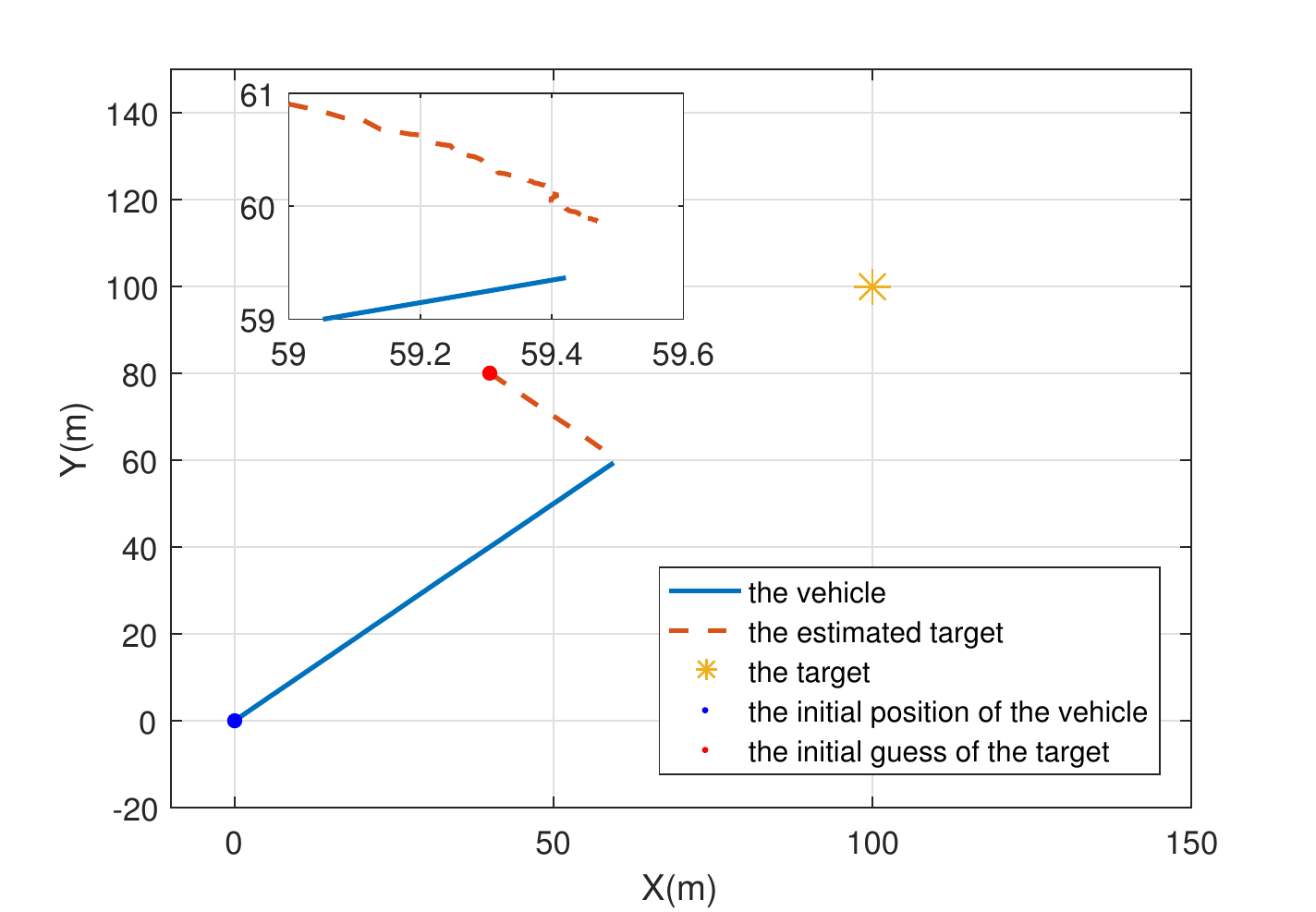} 
\caption{The trajectory of the vehicle and the estimated positions of the target with $\beta=3$.}
\label{fig:tra_beta6}
\end{figure}

In Fig. \ref{fig:e_est}, the time for the estimation error less than $1$m are $t=51$s, $t=64$s, and $t=145$s for the cases of $\beta=1$, $\beta=2$, and $\beta=3$. From Fig. \ref{fig:range}, the time of estimated range in Algorithm \ref{alg:control} less than $1$m  are $t=90$s, $t=71$s, and $t=28$s. That is, the larger the value of $\beta$, the slower the estimation convergence and the faster the vehicle approaching the estimated target position. In particular, we can observe from Fig. \ref{fig:range} that the estimated range easily tends to zero for a large $\beta$, i.e., the vehicle tends to approach the estimated target position. This is consistent with Section \ref{sub:trans_nor} that a larger $\beta$ emphasizes more the importance of the control objective. However, the estimate of the target position may be far from the true target position in the case of a larger $\beta$. Figure \ref{fig:tra_beta6}  further shows that the vehicle approaches a ``wrong'' target due to the large estimation error. Thus, the weighting parameter $\beta$ must be chosen carefully to balance the estimation and the control objectives.

\begin{figure}
  \centering
\includegraphics[width=8cm]{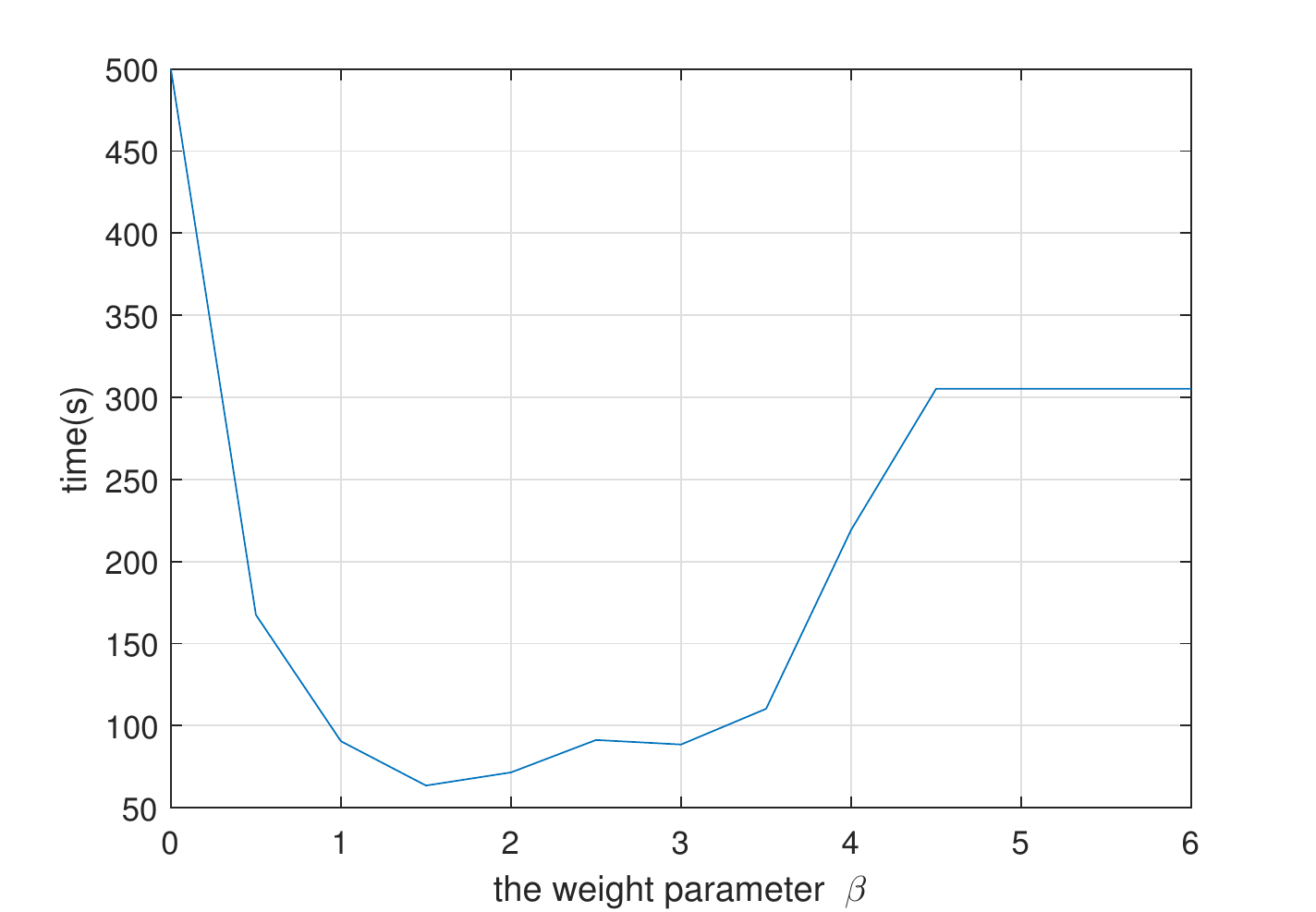} 
\caption{The time for target search with different values of $\beta$.}
\label{fig:beta}
\end{figure}

To further explore {the search performance with different values of $\beta$}, we set a terminal condition of Algorithm \ref{alg:control} as $r(k) < v_c h$, {indicating the completeness of the target search task}. The times for completing the target search task with different values of $\beta$ are given in Fig. \ref{fig:beta}, where the search time initially decreases with a larger $\beta$, and then increases as $\beta$ become much larger. Finally, it keeps constant when $\beta \geqslant 4.5$. This implies that neither too small or too large value of $\beta$ is preferable in the target search problem.

\section{Conclusion}
\label{sec:conclusions}

We have derived an optimization-based controller for the discrete-time Dubins vehicle to address the bearing-only target search problem, where the vehicle was to be steered to the unknown target position as fast as possible and the estimation performance for the unknown target position was considered. In the proposed scheme, we estimated the unknown target position via the recursive estimator with bearing angles in global and the vehicle's coordinate. Using the estimated target position, we formulated and solved a bi-objective optimization problem with the estimation and the control objectives. In particular, the solution was in terms of the predefined importance of the control objective relative to the estimation objective. Then, an optimization-based controller was obtained and drove the vehicle to approach the target as fast as possible while optimizing the estimation performance. Finally, simulations were given to verify the effectiveness of the proposed controller.

\appendices

\section{Normalization of the optimization problem \eqref{eqn:opt2}} 
\label{app:A}

In the appendix, we denote $\dot f(\cdot), \ddot f(\cdot)$ as the derivative and second derivative of the function $f(\cdot)$ with respect to the variable in the bracket.

To normalize the objective function in \eqref{eqn:opt2}, we take the derivative of $f_e(v_r)$ and $f_c(v_r)$ in \eqref{obj_ec} as
\beq \nonumber
\begin{aligned}
\dot f_e(v_r)&= \frac{-2v_r h^2 (r^2-2rv_rh+v_c^2h^2)+4rh^3(v_c^2-v_r^2)}
{\sigma^4r^2(r^2-2rv_rh+v_c^2h^2)^3},\\
\dot f_c(v_r)&= 2rh.
\end{aligned}
\enq
Let $\dot f_e(v_r)=0$, and we obtain a stationary point of $f_e(v_r)$ denoted as $v_s={2rv_c^2h}/({r^2+v_c^2h^2})$. Thus, the increasing interval and decreasing interval of $f_e(v_r)$ are
$[{v_c^2h}/{2r},v_s)$ and $ [v_s,v_c ]$, respectively, and $f_c(v_r)$ is always increasing in the feasible set $[{v_c^2h}/{2r},v_c ]$. It follows that the Pareto-optimal set of the composite optimization problem \eqref{eqn:opt2} is $\left [v_s,v_c\right].$
 
{Instituting $v_r=v_s$ into $f_e(v_r)$ and $f_c(v_r)$}, we obtain the ideal value of $f_e(v_r)$ and the nadir value of $f_c(v_r)$ below
\bee \label{nor1}
f^*_e=\frac{v_c^2h^2}{r^2(r^2-v_c^2h^2)^2},  f_c^{\text{nad}}=\frac{3r^2v_c^2h^2-v_c^4h^4}{r^2+v_c^2h^2}.
\ene
Similarly, taking $v_r=v_c$, we obtain the nadir value of $f_e(v_r)$ and the ideal value of $f_c(v_r)$ below
\bee \label{nor2}
f_e^{\text{nad}}=0, 
f_c^{*}=2rv_ch-v_c^2h^2.
\ene
{Jointly with \eqref{obj_ec}, \eqref{nor1} and \eqref{nor2}}, we can normalize the objective functions as
\beq \nonumber
\begin{aligned}
f_e^{\text{norm}}(v_r)=&\frac{f_e(v_r)-f_e^{\text{nad}}}{f^*_e-f_e^{\text{nad}}}=1-\frac{(v_c^2-v_r^2)(r^2-v_c^2h^2)^2}{v_c^2(r^2-2rv_r h+v_c^2h^2)^2}, \\
f_c^{\text{norm}}(v_r)=&\frac{f_c(v_r)-f_c^{\text{nad}}}{f^*_c-f_c^{\text{nad}}}=\frac{ (v_c-v_r)(v_c^2h^2+r^2)}{v_c(v_ch-r)^2}.
\end{aligned}
\enq 
Then the normalized objective function is given as
$$f(v_r)=f_e^{\text{norm}}(v_r)+ \beta f_c^{\text{norm}}(v_r)$$ with a new weighting parameter $\beta \in[0, +\infty)$.

\section{Proof of Lemma \ref{lemma}} \label{app:B}
The derivative of $f(v_r)$ in (\ref{eqn:opt3}) is computed as
\bee\begin{aligned} \label{eqn:df}
\dot f(v_r)&=\frac{(2v_r(r^2+ v_c^2h^2)-4rv_c^2h)(r^2-v_c^2h^2)^2}{v_c^2(r^2-2rv_r h+v_c^2h^2)^3}\\
&~~~-\beta\frac{v_c^2h^2+r^2}{v_c(v_ch-r)^2}.
\end{aligned}
\ene
Let $\dot f(v_r)=0$, and we have three solutions denoted as $z_1, z_2$, and $z_3$. Since we only focus on the interval $v_r \in [v_s,v_c]$, the first derivatives of $f(v_r)$ at $v_s$ and $v_c$ are computed as follows 
\bee  \label{eqn:dot_fvs}
\dot f(v_s)=0, \dot f(v_c)=\frac{(2-\beta)(v_c^2h^2+r^2)+4rv_c h}{v_c(r+v_ch)^2}.
\ene
Thus, $v_s$ is one of the roots of $\dot f(v_r)=0$, which implies that $\dot f(v_r)=0$ has at least one root in the interval $v_r \in [v_s,v_c]$.

\section{Proof of Theorem \ref{thm:1}} \label{app:C}
Obviously, solutions of the normalized optimization problem (\ref{eqn:opt3}) exist in the Pareto-optimal set, thus we only need to focus on $f(v_r)$ for $v_r \in [v_s,v_c]$. Based on the proof of Lemma \ref{lemma}, {the first derivative of $\dot f(v_c)$ is indefinite, the sign of which is to be analyzed in the following.}

To this end, we firstly consider the continuity of $\dot f(v_r)$ and observe that there exists and only exists a discontinuity point, i.e., $v_r=(r^2+v_c^2h^2)/2rh $, which makes the denominator of $\dot f(v_r)$ equal to zero. Furthermore, this discontinuity point is larger than $v_c$, implying that the function $\dot f(v_r)$ is continuous in the Pareto set $[v_s,v_c]$.

Then, we take the second derivative of $f(v_r)$ in $v_r \in [v_s,v_c]$ and obtain that
\beq \label{eqn:ddf}
\ddot f(v_r)=\frac{2\Theta^2}{v_c^2\Xi^2}+\frac{24r^2h^2\Theta^2(v_r^2-v_c^2)}{v_c^2\Xi^4}+\frac{16rv_rh\Theta^2}{v_c^2\Xi^3},
\enq
where $\Theta=r^2-v_c^2h^2$ and $\Xi=r^2-2v_rhr+v_c^2h^2$. Inserting the bounds of the Pareto-optimal set $v_s, v_c$ into the second derivative yields that
\bee \label{ddot_fvr}
\ddot f(v_s)>0, \ddot f(v_c)>0,
\ene
implying that the function $\dot f(v_r)$ is increasing at $v_s$ and $v_c$.

Now, we focus on the sign of $\dot f(v_c)$. Since the denominator is positive in $[v_s,v_c]$, we only needs to analyze the sign of the numerator $(2-\beta)(v_c^2h^2+r^2)+4rv_c h$.

In light of the fundamental inequality $a^2+b^2 \geqslant 2ab, \forall a, b,$ we obtain that $\beta \geqslant 4$ is a sufficient condition of $\dot f(v_c) \leqslant 0$. The upper subfigure in Fig.\ref{fig:f_df1} approximately depicts $\dot f(v_r)$ in this case, where $\dot f(v_r)$ is continuous in $[v_s,v_c]$, $\dot f(v_s)=0$, $\dot f(v_c) \leqslant 0$, and \eqref{ddot_fvr} holds. Then the function curve of $f(v_r)$ can be approximated corresponding to that of $\dot f(v_r)$, where $ f(v_s)-f(v_c) = \beta-1 > 0$ and $f(v_c)=1$. Therefore, the minimum of $f(v_r)$ is $f(v_c)$.

Similarly, we have that $\beta \leqslant 2$ is a sufficient condition of $\dot f(v_c) \geqslant 0$. There are two cases depicted in the upper subfigures in Fig. \ref{fig:f_df2} and Fig. \ref{fig:f_df3}. In the first case, the other two zero points besides $v_s$ exist in $(v_s, v_c)$, and we denote the largest zero point as $v_z$. Since $f(v_s)-f(v_c) = \beta-1$, it holds that $f(v_z)$ is the minimum if $\beta \geqslant 1$. For $\beta < 1$, the minimum is the smaller one of $f(v_s)$ and $f(v_z)$. Fig. \ref{fig:f_df2} depicts the function curves of $\dot f(v_r)$ and $f(v_r)$, where $\dot f(v_s)=0$, the other two zero points of $\dot f(v_r)$ are in $[v_s, v_c]$, and the corresponding $f(v_r)$ are marked with $\beta \geqslant 1$ and $\beta < 1$, respectively. In the second case, the function $\dot f(v_r)$ remains positive in $(v_s, v_c)$, since there is no zero points. It implies that $f(v_r)$ always increases and $\beta <1$, as Fig. \ref{fig:f_df3} shows. Therefore, the minimum of $f(v_r)$ is $f(v_s)$.

For the case of $\beta \in (2,4)$, it clearly holds that $ f(v_s)-f(v_c) = \beta-1 > 0$. Let $(2-\beta)(v_c^2h^2+r^2)+4rv_c h=0$, and we obtain the following relation that
$$
\rho=\frac{v_ch}{r}=\frac{2-\sqrt{4\beta-\beta^2}}{\beta-2}.
$$
It follows that if $\rho \in \big(0,(2-\sqrt{4\beta-\beta^2})/(\beta-2)\big]$, then $ \dot f(v_c) \leqslant 0$. The function curves are shown in Fig.\ref{fig:f_df1}. Otherwise, the function $\dot f(v_c) > 0$, and the function curves are similar to that in Fig. \ref{fig:f_df2} marked by $\beta \ge 1$.

To summarize, we have following conclusions with different values of $\beta$.

\begin{enumerate}
\renewcommand{\labelenumi}{\rm(\alph{enumi})}
\item \label{a}
If $\beta \in [4,\infty)$, then $\dot f(v_c) \leqslant 0$. It follows from $ f(v_s)-f(v_c) = \beta-1 > 0$ that the minimum of $ f(v_r)$ is $f(v_c)$. The function curves are shown in Fig. \ref{fig:f_df1}.
\begin{figure}
  \centering
\includegraphics[width=8cm]{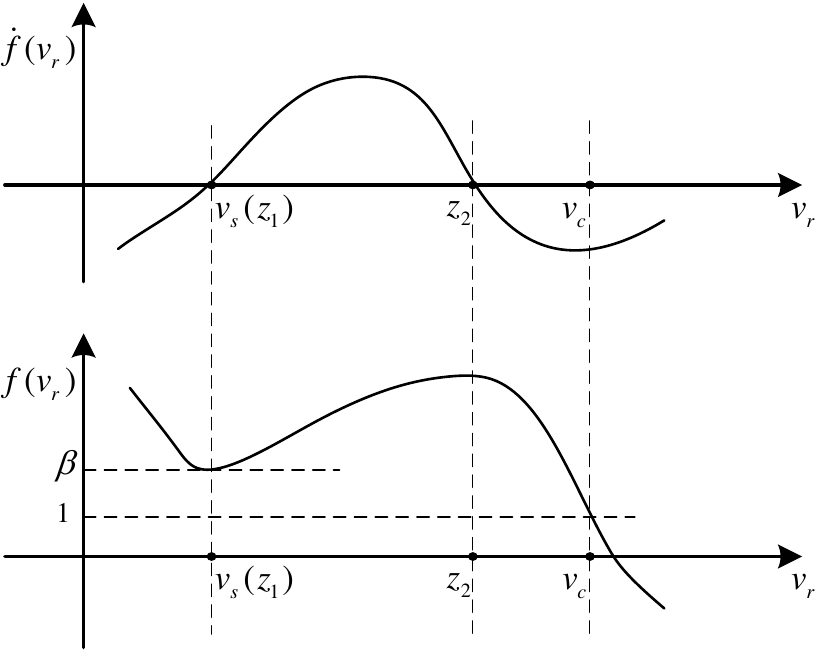} 
\caption{A possible case of $\dot f(v_r)$ and $f(v_r)$ curves.}
\label{fig:f_df1}
\end{figure}

\item \label{b}
If $\beta \in [2,4)$, together with $ f(v_s)-f(v_c) = \beta-1 > 0$, there are two conditions that 

\romannumeral1) If $\rho \in \big(0,(2-\sqrt{4\beta-\beta^2})/(\beta-2)\big]$, then $ \dot f(v_c) \leqslant 0$. Thus, the minimum of $ f(v_r)$ is $f(v_c)$. The function curves are shown in Fig. \ref{fig:f_df1}. 

\romannumeral2) If $\rho \in \big((2-\sqrt{4\beta-\beta^2})/(\beta-2),1 \big)$, then $\dot f(v_c) > 0$. Thus, the minimum of $ f(v_r)$ is $f(v_z)$, where $v_z$ is the largest root of $\dot f(v_r)=0, v_r \in[v_s,v_c]$. The function curves are shown in Fig. \ref{fig:f_df2}.
\begin{figure}
  \centering
\includegraphics[width=8cm]{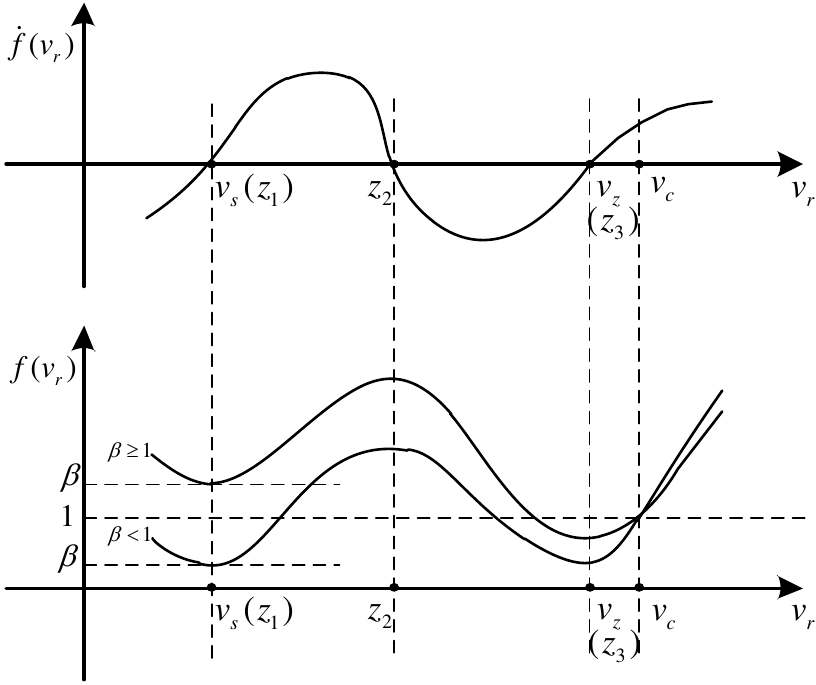} 
\caption{A possible case of $\dot f(v_r)$ and $f(v_r)$ curves.}
\label{fig:f_df2}
\end{figure}

\item \label{b}
If $\beta \in [1,2) $, then $\dot f(v_c) > 0$. It follows from $ f(v_s)-f(v_c) = \beta-1 \geqslant 0$ that the minimum of $ f(v_r)$ is $f(v_z)$. The function curves are shown in Fig. \ref{fig:f_df2}.

\item If $\beta \in [0,1) $, then $\dot f(v_c) > 0$. Together with $ f(v_s)-f(v_c) = \beta-1 < 0$, there are two conditions that\\
\romannumeral1) If there is any root of $\dot f(v_r)=0$ in the interval $(v_s,v_c]$, then the minimum of $ f(v_r)$ is the smaller one of $f(v_z)$ and $f(v_s)$. The function curves are shown in Fig. \ref{fig:f_df2}.\\
\romannumeral2) If there is no root of $\dot f(v_r)=0$ in the interval $(v_s,v_c]$, then the minimum of $ f(v_r)$ is $f(v_s)$. The function curves are shown in Fig. \ref{fig:f_df3}.

\begin{figure}
  \centering
\includegraphics[width=7.5cm]{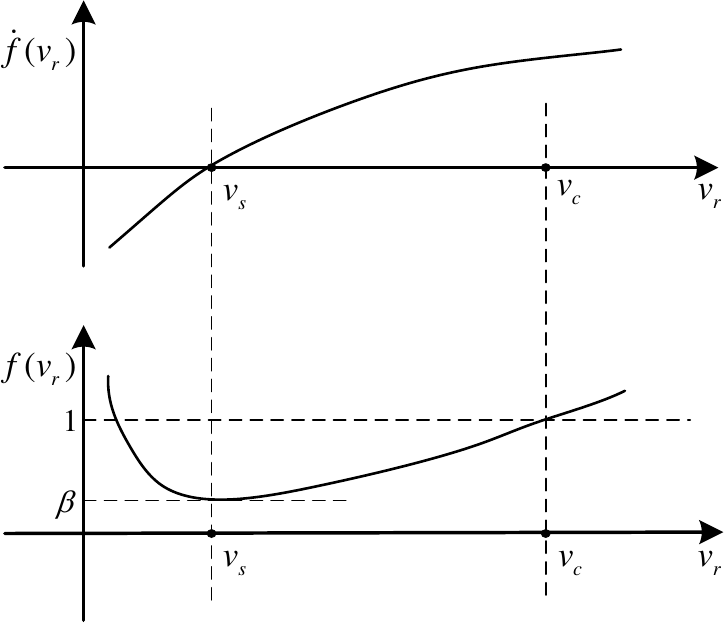} 
\caption{A possible case of $\dot f(v_r)$ and $f(v_r)$ curves.}
\label{fig:f_df3}
\end{figure}
\end{enumerate} 

The optimal solution $v_r^*$ of the normalized optimization problem \eqref{eqn:opt3} can be selected as in Theorem \ref{thm:1}.

\ifCLASSOPTIONcaptionsoff
  \newpage
\fi

\bibliographystyle{IEEETran}
\bibliography{RefDBase}

\begin{IEEEbiography}[{\includegraphics[width=1in,height=1.25in,clip,keepaspectratio]{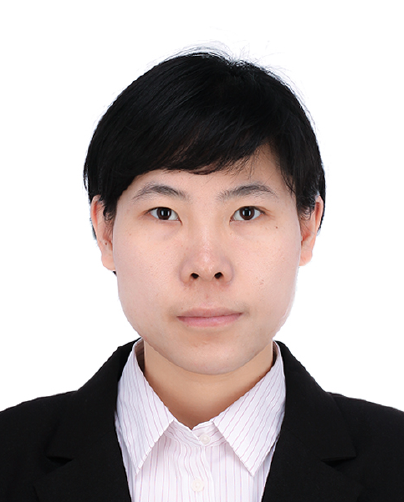}}]{Zhuo Li} received the B.S. degree from the Department of Automation, Harbin Institute of Technology, Harbin, China, in 2016. She is currently pursuing the Ph.D. degree at the Department of Automation, Tsinghua University, Beijing, China. Her  research interests include nonlinear control, distributed optimization, source seeking and learning.
\end{IEEEbiography}

\begin{IEEEbiography}[{\includegraphics[width=1in,height=1.25in,clip,keepaspectratio]{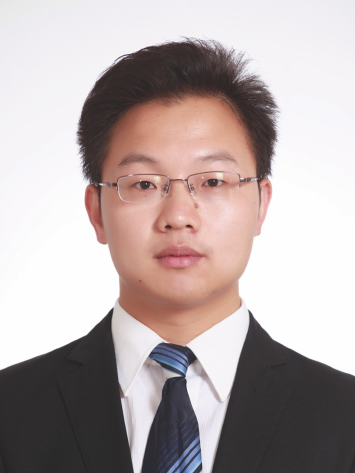}}]
{Keyou You} (SM'17)  received the B.S. degree in Statistical Science from Sun Yat-sen University, Guangzhou, China, in 2007 and the Ph.D. degree in Electrical and Electronic Engineering from Nanyang Technological University (NTU), Singapore, in 2012. After briefly working as a Research Fellow at NTU, he joined Tsinghua University in Beijing, China where he is now a tenured Associate Professor in the Department of Automation. He held visiting positions at Politecnico di Torino, The Hong Kong University of Science and Technology, The University of Melbourne and etc. His current research interests include networked control systems, distributed optimization and learning, and their applications.

	Dr. You received the Guan Zhaozhi award at the 29th Chinese Control Conference in 2010, the CSC-IBM China Faculty Award in 2014 and the ACA (Asian Control Association) Temasek Young Educator Award in 2019. He was selected to the National 1000-Youth Talent Program of China in 2014 and received the National Science Fund for Excellent Young Scholars in 2017. 
	
\end{IEEEbiography}

\begin{IEEEbiography}[{\includegraphics[width=1in,height=1.25in,clip,keepaspectratio]{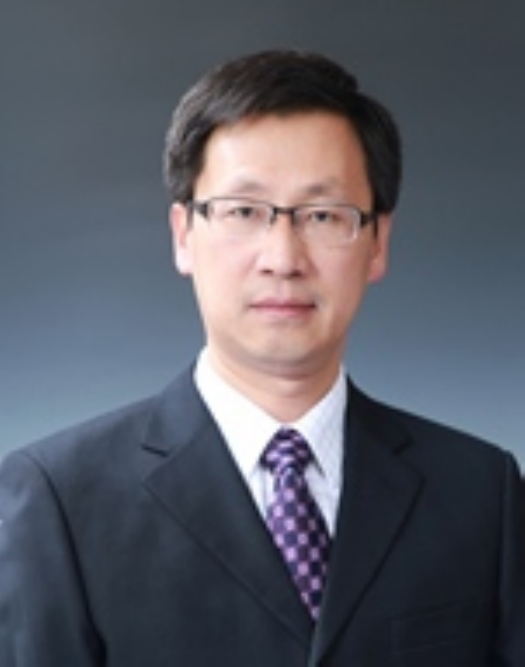}}]{Shiji Song}
received the Ph.D. degree in the Department of Mathematics from Harbin Institute of Technology in 1996. He is a professor in the Department of Automation, Tsinghua University. His research interests include system modeling, control and optimization, computational intelligence and pattern recognition.
\end{IEEEbiography}

\begin{IEEEbiography}[{\includegraphics[width=1in,height=1.25in,clip,keepaspectratio]{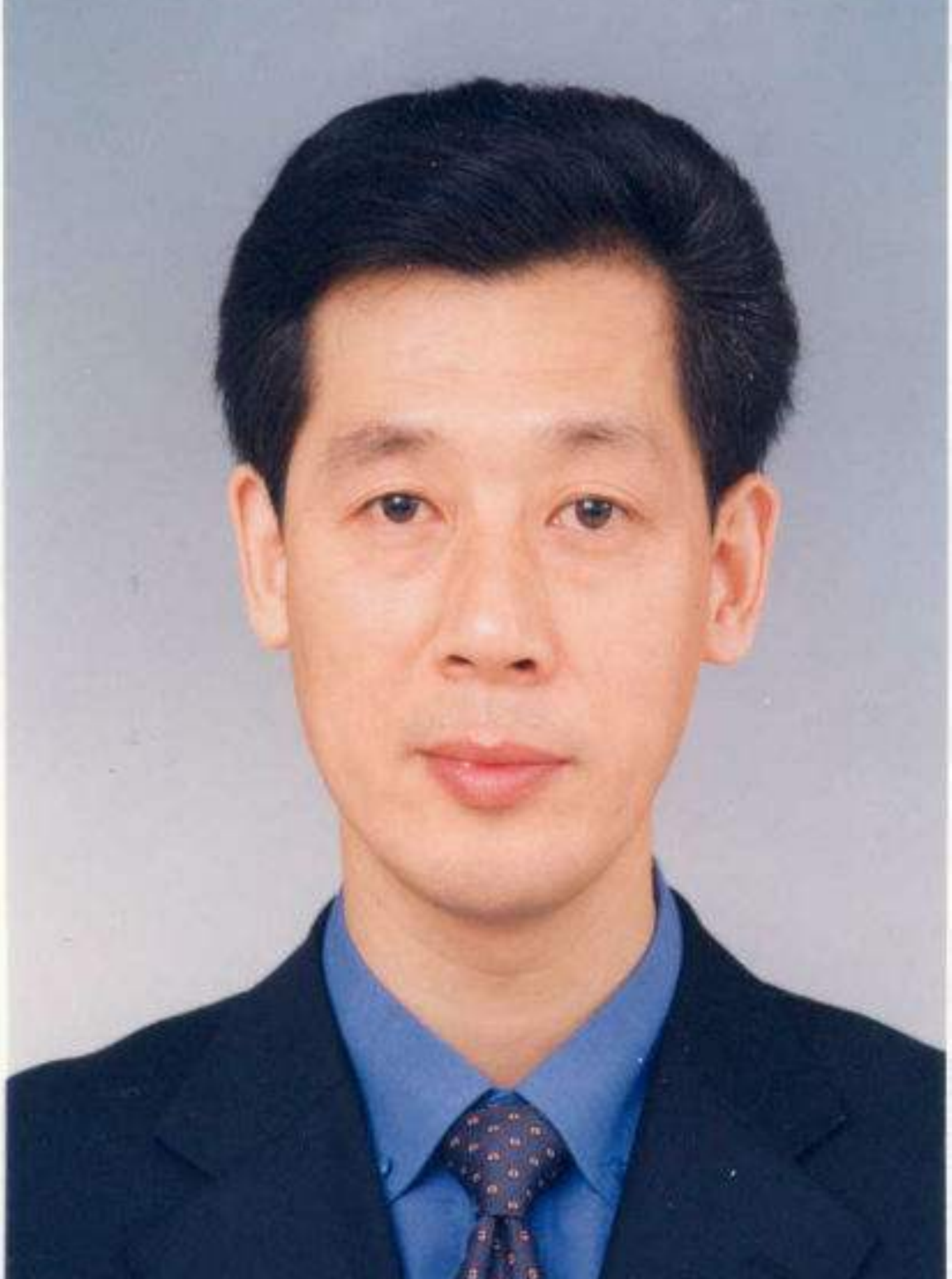}}]{Anke Xue} received the Ph.D. degree from Zhejiang University, in 1997. He is a professor in Hangzhou Dianzi University. His current research interests include robust control theory and applications.
\end{IEEEbiography}

\end{document}